\documentclass[10pt,journal,final,letterpaper,twoside,twocolumn]{IEEEtran}

\usepackage{algorithm}
\usepackage{algorithmic}
\usepackage{graphicx}
\usepackage[cmex10]{amsmath}
\usepackage{amsfonts}
\usepackage{amssymb}
\usepackage{subfigure}
\usepackage{color}
\usepackage[all,poly]{xy}
\usepackage{multirow}
\usepackage[noadjust]{cite}
\usepackage{verbatim}
\usepackage{dsfont}
\usepackage{euscript}

\usepackage{epic,eepic,pstricks}
\usepackage[numbers, square, comma, sort&compress]{natbib}
\usepackage{subeqnarray}
\usepackage{lipsum}
\usepackage{relsize}

\usepackage{enumerate}
\usepackage{theorem}
\usepackage{pifont}   
\usepackage{graphicx}

\graphicspath{{figures/}}

\newcommand{\bx}{\boldsymbol{x}}
\newcommand{\bw}{\boldsymbol{w}}
\newcommand{\bu}{\boldsymbol{u}}
\newcommand{\bv}{\boldsymbol{v}}
\newcommand{\by}{\boldsymbol{y}}
\newcommand{\bz}{\boldsymbol{z}}
\newcommand{\bh}{\boldsymbol{h}}
\newcommand{\btheta}{\boldsymbol{\theta}}

\newcommand{\argmin}{\operatornamewithlimits{argmin}}
\newcommand{\bH}{\boldsymbol{H}}
\newcommand{\bD}{\boldsymbol{D}}
\newcommand{\RX}{\ensuremath{\left]-\infty,+\infty\right]}}
\newcommand{\minimize}[2]{\ensuremath{\underset{\substack{{#1}}}%
{\text{\rm minimize}}\;#2 }}
\newcommand{\E}{{\mathsf E}} 
\newcommand{\RR}{\ensuremath{\mathbb{R}}}
\newcommand{\XX}{\ensuremath{\boldsymbol{\EuScript{X}}}}
 
\newcommand{\prox}{\ensuremath{\operatorname{prox}}}

\newcommand{\ModifPS}[1]{\textcolor{black}{#1}}
\newcommand{\ModifJCP}[1]{\textcolor{black}{#1}}

\newcommand*\dif{\mathop{}\!\mathrm{d}} 
\DeclareMathOperator{\Ex}{E} 
\DeclareMathOperator{\var}{var} 
\newcommand{\Real}{\mathbb{R}} 
\newcommand{\defn}{\triangleq} 
\newcommand{\Normpdf}{\mathcal{N}} 
\newcommand{\Dkl}{D} 
\usepackage{pstool} 

\title{A  Survey of Stochastic Simulation and Optimization Methods in Signal Processing}
\author{Marcelo Pereyra, Philip Schniter, Emilie Chouzenoux, Jean-Christophe Pesquet,\\  Jean-Yves Tourneret, Alfred Hero, and Steve McLaughlin 

\thanks{This work was funded in part by the SuSTaIN program - EPSRC grant EP/D063485/1 - at the Department of Mathematics, University of Bristol. 
SMcL acknowledges the support of EPSRC via grant EP/J015180/1.
MP holds a Marie Curie Intra-European Fellowship for Career Development.
PS acknowledges the support of the NSF via grants CCF-1218754 and CCF-1018368. AH acknowledges the support of ARO grant W911NF-15-1-0479.  Part of this work was also supported by the CNRS Imag'in project under grant 2015OPTIMISME, by the HYPANEMA ANR Project under Grant ANR-12-BS03-003, by the BNPSI ANR Project no ANR-13- BS-03-0006-01, and by the project ANR-11-L ABX-0040-CIMI as part of the program ANR-11-IDEX-0002-02 within the thematic trimester on image processing.}

\thanks{Marcelo Pereyra is with the University of Bristol, School of Mathematics, University Walk, BS8 1TW, UK (e-mail: marcelo.pereyra@bristol.ac.uk).}
\thanks{Philip Schniter is with The Ohio State University, Department of Electrical and Computer Engineering, 2015 Neil Ave., Columbus, OH 43210, USA (e-mail: schniter@ece.osu.edu).}
\thanks{Emilie Chouzenoux and Jean-Christophe Pesquet are with the Laboratoire d'Informatique Gaspard Monge, Universit\'e Paris-Est, Champs sur Marne, France (email: 
\{emilie.chouzenoux,jean-christophe.pesquet\}@univ-paris-est)}
\thanks{Jean-Yves Tourneret is with the University of Toulouse, INP-ENSEEIHT-IRIT-TeSA, France (email: jean-yves.tourneret@enseeiht.fr)}
\thanks{Alfred Hero is with University of Michigan, Dept of Electrical Engineering and Computer Science, USA (email: hero@eecs.umich.edu)}
\thanks{Steve McLaughlin is with Heriot Watt University, Engineering and Physical Sciences, Edinburgh, EH14 4AS, UK (e-mail: s.mclaughlin@hw.ac.uk).}
}

\begin{document}
\setlength{\arraycolsep}{0.5mm}

\maketitle

\begin{abstract}
Modern signal processing (SP) methods rely very heavily on probability and statistics to solve challenging SP problems. SP methods are now expected to deal with ever more complex models, requiring ever more sophisticated computational inference techniques. This has driven the development of statistical SP methods based on stochastic simulation and optimization. Stochastic simulation and optimization algorithms are computationally intensive tools for performing statistical inference in models that are analytically intractable and beyond the scope of deterministic inference methods. They have been recently successfully applied to many difficult problems involving complex statistical models and sophisticated (often Bayesian) statistical inference techniques. This survey paper offers an introduction to stochastic simulation and optimization methods in signal and image processing. The paper addresses a variety of high-dimensional Markov chain Monte Carlo (MCMC) methods as well as deterministic surrogate methods, such as variational Bayes, the Bethe approach, belief and expectation propagation and approximate message passing algorithms. It also discusses a range of optimization methods that have been adopted to solve stochastic problems, as well as stochastic methods for deterministic optimization. Subsequently, areas of overlap between simulation and optimization, in particular optimization-within-MCMC and MCMC-driven optimization are discussed. 
\end{abstract}


\section{Introduction}

Modern signal processing (SP) methods, (we use SP here to cover all relevant signal and image processing problems), rely very heavily on probabilistic and statistical tools; for example, they use stochastic models to represent the data observation process and the prior knowledge available, and they obtain solutions by performing statistical inference (e.g., using maximum likelihood or Bayesian strategies). Statistical SP methods are, in particular, routinely applied to many and varied tasks and signal modalities, ranging from resolution enhancement of medical images to hyperspectral image unmixing; from user rating prediction to change detection in social networks; and from source separation in music analysis to speech recognition.

However, the expectations and demands on the performance of such methods are constantly rising. SP methods are now expected to deal with challenging problems that require ever more complex models, and more importantly, ever more sophisticated computational inference techniques. This has driven the development of computationally intensive SP methods based on stochastic simulation and optimization. 
Stochastic simulation and optimization algorithms are computationally intensive tools for performing statistical inference in models that are analytically intractable and beyond the scope of deterministic inference methods. They have been recently successfully applied to many difficult SP problems involving complex statistical models and sophisticated (often Bayesian) statistical inference analyses. These problems can generally be formulated as inverse problems involving partially unknown observation processes and imprecise prior knowledge, for which they delivered accurate and insightful results. These stochastic algorithms are also closely related to the randomized, variational Bayes and message passing algorithms that are pushing forward the state of the art in approximate statistical inference for very large-scale problems. \black{The key thread that makes stochastic simulation and optimization methods appropriate for these applications is the complexity and high dimensionality involved. For example in the case of hyperspectral imaging the data being processed \black{can involve} \black images of 2048 by 1024 pixels across up to \black{hundreds or thousands of} \black wavelengths.} \black

 \black{This survey paper offers an introduction to stochastic simulation and optimization methods in signal and image processing. The paper addresses a variety of high-dimensional Markov chain Monte Carlo (MCMC) methods as well as deterministic surrogate methods, such as variational Bayes, the Bethe approach, belief and expectation propagation and approximate message passing algorithms. It also discusses a range of stochastic optimization approaches. Subsequently, areas of overlap between simulation and optimization, in particular optimization-within-MCMC and MCMC-driven optimization are discussed. \black{Some methods such as sequential Monte Carlo methods or methods based on importance sampling are not considered in this survey mainly due to space limitations.}\black

This paper seeks to provide a survey of a variety of the algorithmic approaches in a tutorial fashion, as well as to highlight the state of the art, relationships between the methods, and potential future directions of research. In order to set the scene and inform our notation, consider an unknown random vector of interest $\bx=[x_1,\dots,x_N]^T$ and an observed data vector $\by=[y_1,\dots,y_M]^T$, related to $\bx$ by a statistical model with likelihood function $p(\by|\bx;\btheta)$ potentially parametrized by a deterministic vector of parameters $\btheta$. Following a Bayesian inference approach, we model our prior knowledge about $\bx$ with a prior distribution $p(\bx;\btheta)$, and our knowledge about $\bx$ after observing $\by$ with the posterior distribution
\begin{equation}\label{eq:posterior}
p(\bx|\by;\btheta) 
= \frac{ p(\by|\bx;\btheta) p(\bx;\btheta) 
        }{ Z(\btheta) } 
\end{equation}
where the normalising constant
\begin{equation}\label{eq:partition}
Z(\btheta)
= p(\by;\btheta) 
= \int p(\by|\bx;\btheta) p(\bx;\btheta) \dif\bx
\end{equation}
is known as the ``evidence'', model likelihood, or the partition function.
Although the integral in \eqref{eq:partition} suggests that all $x_j$ are continuous random variables, we allow any random variable $x_j$ to be either continuous or discrete, and replace the integral with a sum as required.

In many applications, we would like to evaluate the posterior $p(\bx|\by;\btheta)$ or some summary of it,
for instance point estimates of $\bx$ such as the conditional mean (i.e., MMSE estimate) $\Ex\{\bx|\by;\btheta\}$, uncertainty reports such as the conditional variance $\var\{\bx|\by;\btheta\}$, or expected log statistics as used in the expectation maximization (EM) algorithm \cite{Dempster:JRSS:77,Neal:Jordan:98,Attias:NIPS:00}
\begin{align}
\btheta^{(i+1)}
&= \arg\max_{\btheta} \Ex\{\ln p(\bx,\by;\btheta)|\by;\btheta^{(i)}\}
\end{align}
\black{where the expectation is taken with respect to $p(\bx | \by;\btheta^{(i)})$.} \black However, when the signal dimensionality $N$ is large, the integral in \eqref{eq:partition}, as well as those used in the posterior summaries, are often computationally intractable.
Hence, the interest in computationally efficient alternatives. 
\black{
An alternative that has received a lot of attention in the statistical SP community is maximum-a-posteriori (MAP) estimation.
Unlike other posterior summaries, MAP estimates can be computed by finding the value of $\bx$ maximizing $p(\bx|\by;\btheta) $, which for many models is 
significantly more computationally tractable than numerical integration.} \black In the sequel, we will suppress the dependence on $\btheta$ in the notation, since it is not of primary concern.

The paper is organized as follows. After this brief introduction where we have introduced the basic notation adopted, Section \ref{Sec2} discusses stochastic simulation methods, and in particular a variety of MCMC methods. In Section \ref{Sec3} we discuss deterministic surrogate methods, such as variational Bayes, the Bethe approach, belief and expectation propagation, and provide a brief summary of approximate message passing algorithms. Section \ref{Sec4} discusses a range of optimization methods for solving stochastic problems, as well as stochastic methods for solving deterministic optimization problems. Subsequently, in Section \ref{Sec5} we discuss areas of overlap between simulation and optimization, in particular the use of optimization techniques within MCMC algorithms and MCMC-driven optimization, and suggest some interesting areas worthy of exploration. Finally, in Section \ref{Sec6} we draw together thoughts, observations and conclusions.

\section{Stochastic Simulation methods}\label{Sec2}
Stochastic simulation methods are sophisticated random number generators that allow samples to be drawn from a user-specified target density $\pi(\bx)$, such as the posterior $p(\bx|\by)$. These samples can then be used, for example, to approximate probabilities and expectations by Monte Carlo integration \cite[Ch. 3]{casella:robert:2004}. In this section we will focus on Markov chain Monte Carlo (MCMC) methods, an important class of stochastic simulation techniques that operate by constructing a Markov chain with stationary distribution $\pi$. In particular, we concentrate on Metropolis-Hastings algorithms for high-dimensional models (see \cite{green:2015} for a more general recent review of MCMC methods). It is worth emphasizing, however, that we do not discuss many other important approaches to simulation that also arise often in signal processing applications, such as ``particle filters'' or sequential Monte Carlo methods \cite{doucet:2013,BeskosJMS}, and approximate Bayesian computation \cite{marin:pudlo:robert:ryder:2011}.

A cornerstone of MCMC methodology is the Metropolis-Hastings (MH) algorithm \cite[Ch. 7]{casella:robert:2004}\cite{metropolis:1953, hastings:1970}, a universal machine for constructing Markov chains with stationary density $\pi$. Given some generic proposal distribution $\bx^{*} \sim q(\cdot|\bx)$, the generic MH algorithm proceeds as follows.

\begin{algorithm}[H]
\caption{Metropolis--Hastings algorithm (generic version)}
\begin{algorithmic}
\STATE Set an initial state $\bx^{(0)}$
\FOR {$t=1$ to $T$}
\STATE Generate a candidate state $\bx^{*}$ from a proposal $q (\cdot|\bx^{(t-1)})$
\STATE Compute the acceptance probability 
$$
\rho^{(t)}= \min\left(1, \frac{\pi(\bx^{*})}{\pi(\bx^{(t-1)})}\frac{q (\bx^{(t-1)}|\bx^{*})}{
q(\bx^{*}|\bx^{(t-1)})}\right)
$$
\STATE Generate $u_t\sim\mathcal{U}(0,1)$
\IF{$u_t\le\rho^{(t)}$}
\STATE Accept the candidate and set $\bx^{(t)}=\bx^{*}$
\ELSE
\STATE Reject the candidate and set $\bx^{(t)}=\bx^{(t-1)}$
\ENDIF
\ENDFOR
\end{algorithmic}
\label{algo:MH.0}
\end{algorithm}
Under mild conditions on $q$, the chains generated by Algo. \ref{algo:MH.0} are ergodic and converge to the stationary distribution $\pi$ \cite{chib:greenberg:1995,Mylene:2007}. An important feature of this algorithm is that computing the acceptance probabilities $\rho^{(t)}$ does not require knowledge of the normalization constant of $\pi$ (which is often not available in Bayesian inference problems). \black{The intuition for the MH algorithm is that the algorithm proposes a stochastic perturbation to the state of the chain and applies a carefully defined decision rule to decide if this perturbation should be accepted or not. This decision rule, given by the random accept-reject step in Algo. \ref{algo:MH.0}, ensures that at equilibrium the samples of the Markov chain have $\pi$ as marginal distribution.}\black

The specific choice of $q$ will of course determine the efficiency and the convergence properties of the algorithm. Ideally one should choose $q = \pi$ to obtain a perfect sampler (i.e., with candidates accepted with probability $1$); this is of course not practically feasible since the objective is to avoid the complexity of directly simulating from $\pi$. In the remainder of this section we review strategies for specifying $q$ for high-dimensional models, and discuss relative advantages and disadvantages. \black{In order to compare and optimize the choice of $q$, a performance criterion needs to be chosen. A natural criterion is the stationary integrated autocorrelation time for some relevant scalar summary statistic $g : \mathbb{R}^{\black{N}} \rightarrow \mathbb{R}$, i.e.,
\begin{equation}\label{siat}
\tau_g = 1 + 2\sum_{t = 1}^\infty \textrm{Cor}\{g(\bx^{(0)}),g(\bx^{(t)})\}
\end{equation}
\black{with $\bx^{(0)} \sim \pi$, and where $\textrm{Cor}(\cdot,\cdot)$ denotes the correlation operator}. This criterion is directly related to the effective number of independent Monte Carlo samples produced by the MH algorithm, and therefore to the mean square error of the resulting Monte Carlo estimates. Unfortunately drawing conclusions directly from \eqref{siat} is generally not possible because $\tau_g$ is highly dependent on the choice of $g$, with different choices often leading to contradictory results. Instead, MH algorithms are generally studied asymptotically in the infinite-dimensional model limit. More precisely, in the limit $N\rightarrow \infty$, the algorithms can be studied using diffusion process theory, where the dependence on $g$ vanishes and all measures of efficiency become proportional to the diffusion speed. The ``complexity'' of the algorithms can then be defined as the rate at which efficiency deteriorates as $N \rightarrow \infty$, \black{e.g.,} \black $\mathcal{O}(N)$ (see \cite{roberts:2001} for an introduction to this topic and details about the relationship between the efficiency of MH algorithms and their average acceptance probabilities or acceptance rates)\footnote{Notice that this measure of complexity of MCMC algorithms does not take into account the computational costs related to generating candidate states and evaluating their Metropolis-Hastings acceptance probabilities, which typically also scale at least linearly with the problem dimension $N$.}.}\black

Finally, it is worth mentioning that despite the generality of this approach, there are some specific models for which conventional MH sampling is not possible because the computation of $\rho^{(t)}$ is intractable (e.g., when $\pi$ involves an intractable function of $\bx$, such as the partition function of the Potts-Markov random field). This issue has received a lot of attention in the recent MCMC literature, and there are now several variations of the MH construction for intractable models \cite{andrieu:2009, andrieu:2010,marin:pudlo:robert:ryder:2011, Pereyra_IEEE_Trans_IP_2013}.

\subsection{Random walk Metropolis-Hastings algorithms}
The so-called \emph{random walk Metropolis-Hastings} (RWMH) algorithm is based on proposals of the form $\bx^* = \bx^{(t-1)} + \bw$, where typically 
$\bw \sim \mathcal{N}(\boldsymbol{0},\boldsymbol{\Sigma})$ for some positive-definite covariance matrix $\boldsymbol{\Sigma}$ \cite[Ch. 7.5]{casella:robert:2004}. This algorithm is one of the most widely used MCMC methods, perhaps because it has very robust convergence properties and a relatively low computational cost per iteration. It can be shown that the RWMH algorithm is geometrically ergodic under mild conditions on $\pi$ \cite{jarner2000geometric}. Geometric ergodicity is important because it guarantees a central limit theorem for the chains, and therefore that the samples can be used for Monte Carlo integration. However, the myopic nature of the random walk proposal means that the algorithm often requires a large number of iterations to explore the parameter space, and will tend to generate samples that are highly correlated, particularly if the dimension $N$ is large (the performance of this algorithm generally deteriorates at rate $\mathcal{O}(N)$, which is worse than other more advanced stochastic simulation MH algorithms \cite{Beskos:2009}). This drawback can be partially mitigated by adjusting the proposal matrix $\boldsymbol{\Sigma}$ to approximate the covariance structure of $\pi$, and some adaptive versions of RWHM perform this adaptation automatically. For sufficiently smooth target densities, performance is further optimized by scaling $\boldsymbol{\Sigma}$ to achieve an acceptance probability of approximately $20\%-25\%$ \cite{Beskos:2009}.
 
\subsection{Metropolis adjusted Langevin algorithms}
The Metropolis adjusted Langevin algorithm (MALA) is an advanced MH algorithm inspired by the Langevin diffusion process on $\mathbb{R}^N$, defined as the solution to the stochastic differential equation \cite{roberts:1996}
\begin{equation}\label{diffusion}
dX(t) = \frac{1}{2}\nabla\log\pi\left(X(t)\right)dt + dW(t), \quad X(0) = \bx_0
\end{equation}
where $W$ is the Brownian motion process on $\mathbb{R}^N$ and $\bx_0 \in \mathbb{R}^N$ denotes some initial condition. Under appropriate stability conditions, $X(t)$ converges in distribution to $\pi$ as $t \rightarrow \infty$, and is therefore potentially interesting for drawing samples from $\pi$. Since direct simulation from $X(t)$ is only possible in very specific cases, we consider a discrete-time forward Euler approximation to \eqref{diffusion} given by
\begin{equation}\label{ULA}
X^{(t+1)} \sim \mathcal{N}\left(X^{(t)} + \frac{\delta}{2}\nabla\log\pi\left(X^{(t)}\right), \delta\mathbb{I}_N\right)
\end{equation}
where the parameter $\delta$ controls the discrete-time increment. Under certain conditions on $\pi$ and $\delta$, \eqref{ULA} produces a good approximation of $X(t)$ and converges to a stationary density which is close to $\pi$. In MALA this approximation error is corrected by introducing an MH accept-reject step that guarantees convergence to the correct target density $\pi$. The resulting algorithm is equivalent to Algo. \ref{algo:MH.0} above, with proposal 
\begin{equation}\label{propMALA}
\begin{split}
q&(\bx^*|\bx^{(t-1)}) = \\
&\frac{1}{(2\pi\delta)^{N/2}}\exp{\left(-\frac{\|\bx^* - \bx^{(t-1)} - \frac{\delta}{2}\nabla\log\pi\left(\bx^{(t-1)}\right) \|^2}{2\delta}\right)}.
\end{split}
\end{equation}
\black{\black{By analyzing} \black the proposal \eqref{propMALA} we notice that, in addition to the Langevin interpretation, MALA can also be interpreted as an MH algorithm that, at each iteration $t$, draws a candidate from a local quadratic approximation to $\log\pi$ around $\bx^{(t-1)}$, with $\delta^{-1}\mathbb{I}_N$ as an approximation to the Hessian matrix}.\black

In addition, the MALA proposal can also be defined using a matrix-valued time step $\boldsymbol{\Sigma}$. This modification is related to redefining \eqref{ULA} in an Euclidean space with the inner product $\langle \bw, \boldsymbol{\Sigma}^{-1}\bx\rangle$ \cite{girolami:2011}. \black{Again, the matrix $\boldsymbol{\Sigma}$ should capture the correlation structure of $\pi$ to improve efficiency. For example, $\boldsymbol{\Sigma}$ can be the spectrally-positive version of the inverse Hessian matrix of $\log\pi$ \cite{betancourt:2013}, or the inverse Fisher information matrix \black{of the statistical observation model} \black \cite{girolami:2011}. \black{Note that}, \black in a similar fashion to preconditioning in optimization, using the exact full Hessian or Fisher information matrix is often too computationally expensive in high-dimensional settings and more efficient representations must be used instead}\black. Alternatively, adaptive versions of MALA can learn a representation of the covariance structure online \cite{Atchade2006}. For sufficiently smooth target densities, MALA's performance can be further optimized by scaling $\boldsymbol{\Sigma}$ (or $\delta$) to achieve an acceptance probability of approximately $50\% - 60\%$ \cite{pillai2012optimal}.

Finally, there has been significant empirical evidence that MALA can be very efficient for some models, particularly in high-dimensional settings and when the cost of computing the gradient $\nabla\log\pi(\bx)$ is low. Theoretically, for sufficiently smooth $\pi$, the complexity of MALA scales at rate $\mathcal{O}(N^{1/3})$ \cite{pillai2012optimal}, comparing very favorably to the $\mathcal{O}(N)$ rate of RWMH algorithms. However, the convergence properties of the conventional MALA are not as robust as those of the RWMH algorithm. In particular, MALA may fail to converge if the tails of $\pi$ are super-Gaussian or heavy-tailed, or if $\delta$ is chosen too large \cite{roberts:1996}. Similarly, MALA might also perform poorly if $\pi$ is not sufficiently smooth, or multi-modal. Improving MALA's convergence properties is an active research topic. \black{Many limitations of the original MALA algorithm} \black can now be avoided by using more advanced versions \cite{pereyra:2015,xifara:2014,girolami:2011,casella2011stability,schreck2013shrinkage}.

\subsection{Hamiltonian Monte Carlo}
The Hamiltonian Monte Carlo (HMC) method is a very elegant and successful instance of an MH algorithm based on auxiliary variables \cite{neal:2013}. Let $\bw \in \mathbb{R}^N$, $\boldsymbol{\Sigma} \in \mathbb{R}^{N\times N}$ positive definite, and consider the augmented target density $\pi(\bx,\bw) \propto \pi(\bx)\exp(-\frac{1}{2}\bw^T \boldsymbol{\Sigma}^{-1} \bw)$, which admits the desired target density $\pi(\bx)$ as marginal. The HMC method is based on the observation that the trajectories defined by the so-called \emph{Hamiltonian dynamics} preserve the level sets of $\pi(\bx,\bw)$. A point $(\bx_0,\bw_0) \in \mathbb{R}^{2N}$ that evolves according to the differential equations \eqref{Hamiltonian} during some simulation time period $(0,t]$
\begin{equation}\label{Hamiltonian}
\begin{split}
\frac{\textrm{d}\bx}{\textrm{d}t} &= - \nabla_{\bw} \log \pi(\bx,\bw) = \boldsymbol{\Sigma}^{-1}\bw\\
\frac{\textrm{d}\bw}{\textrm{d}t} &= \nabla_{\bx} \log \pi(\bx,\bw) = \nabla_{\bx} \log \pi(\bx)
\end{split}
\end{equation}
yields a point $(\bx_t, \bw_t)$ that verifies $\pi(\bx_t,\bw_t) = \pi(\bx_0,\bw_0)$. In MCMC terms, the deterministic proposal \eqref{Hamiltonian} has $\pi(\bx,\bw)$ as invariant distribution. Exploiting this property for stochastic simulation, the HMC algorithm combines \eqref{Hamiltonian} with a stochastic sampling step, $\bw \sim \mathcal{N}(0,\boldsymbol{\Sigma})$, that also has invariant distribution $\pi(\bx,\bw)$, and that will produce an ergodic chain. Finally, as with the Langevin diffusion \eqref{diffusion}, it is generally not possible to solve the Hamiltonian equations \eqref{Hamiltonian} exactly. Instead, we use a \emph{leap-frog} approximation detailed in \cite{neal:2013}
\begin{eqnarray}\nonumber
&\bw^{(t+\delta/2)} &= \bw^{(t)} + \frac{\delta}{2}\nabla_{\bx} \log \pi\left(\bx^{(t)}\right) \\\label{Leapfrog}
&\bx^{(t+\delta)} &= \bx^{(t)} + \delta\boldsymbol{\Sigma}^{-1}\bw^{(t+\delta/2)}\\\nonumber
&\bw^{(t+\delta)} &= \bw^{(t+\delta/2)} + \frac{\delta}{2}\nabla_{\bx} \log \pi\left(\bx^{(t+\delta)}\right)
\end{eqnarray}
where again the parameter $\delta$ controls the discrete-time increment. The approximation error introduced by \eqref{Leapfrog} is then corrected with an MH step targeting $\pi(\bx,\bw)$. This algorithm is summarized in Algo. \ref{algo:HMC} below \black{(see \cite{neal:2013} for details about the derivation of the acceptance probability)}\black.

\begin{algorithm}[H]
\caption{Hamiltonian Monte Carlo (with leap-frog)}
\begin{algorithmic}
\STATE Set an initial state $\bx^{(0)}$, $\delta > 0$, and $L \in \mathbb{N}$.
\FOR {$t=1$ to $T$}
\STATE Refresh the auxiliary variable $\bw \sim \mathcal{N}(0,\boldsymbol{\Sigma})$.
\STATE Generate a candidate $(\bx^*,\bw^*)$ by propagating the current state $(\bx^{(t-1)},\bw)$ with $L$ leap-frog steps of length $\delta$ defined in \eqref{Leapfrog}. 
\STATE Compute the acceptance probability 
$$
\rho^{(t)}= \min\left(1, \frac{\pi(\bx^{*},\bw^*)}{\pi(\bx^{(t-1)},\bw)}\right).
$$
\STATE Generate $u_t\sim\mathcal{U}(0,1)$.
\IF{$u_t\le\rho^{(t)}$}
\STATE Accept the candidate and set $\bx^{(t)}=\bx^{*}$.
\ELSE
\STATE Reject the candidate and set $\bx^{(t)}=\bx^{(t-1)}$.
\ENDIF
\ENDFOR
\end{algorithmic}
\label{algo:HMC}
\end{algorithm}

Note that to obtain samples from the marginal $\pi(\bx)$ it is sufficient to project the augmented samples $(\bx^{(t)},\bw^{(t)})$ onto the original space of $\bx$ (i.e., by discarding the auxiliary variables $\bw^{(t)}$). It is also worth mentioning that under some regularity condition on $\pi(\bx)$, the leap-frog approximation \eqref{Leapfrog} is time-reversible and volume-preserving, and that these properties are key to the validity of the HMC algorithm \cite{neal:2013}. 

Finally, there has been a large body of empirical evidence supporting HMC, particularly for high-dimensional models. Unfortunately, its theoretical convergence properties are much less well understood \cite{betancourt:2014}. It has been recently established that for certain target densities the complexity of HMC scales at rate $\mathcal{O}(N^{1/4})$, comparing favorably with MALA's rate $\mathcal{O}(N^{1/3})$ \cite{beskos:2013optimal}. However, it has also been observed that, as with MALA, HMC may fail to converge if the tails of $\pi$ are super-Gaussian or heavy-tailed, or if $\delta$ is chosen too large. HMC may also perform poorly if $\pi$ is multi-modal, or not sufficiently smooth. 

Of course, the practical performance of HMC also depends strongly on the algorithm parameters $\boldsymbol{\Sigma}$, $L$ and $\delta$ \cite{betancourt:2014}. The covariance matrix $\boldsymbol{\Sigma}$ should be designed to model the correlation structure of $\pi(\bx)$, which can be determined by performing pilot runs, or alternatively by using the strategies described in \cite{girolami:2011,betancourt:2013,zhang:2011}. The parameters $\delta$ and $L$ should be adjusted to obtain an average acceptance probability of approximately $60\% - 70\%$ \cite{beskos:2013optimal}. Again, this can be achieved by performing pilot runs, or by using an adaptive HMC algorithm that adjusts these parameters automatically \cite{hoffman:2014,betancourt:2014optimizing}.

\subsection{Gibbs sampling}
The Gibbs sampler (GS) is another very widely used MH algorithm which operates by updating the elements of $\bx$ individually, or by groups, using the appropriate conditional distributions \cite[Ch. 10]{casella:robert:2004}. This divide-and-conquer strategy often leads to important efficiency gains, particularly if the conditional densities involved are ``simple'', in the sense that it is computationally straightforward to draw samples from them. For illustration, suppose that we split the elements of $\bx$ in three groups $\bx = (\bx_1, \bx_2, \bx_3)$, and that by doing so we obtain three conditional densities $\pi(\bx_1|\bx_2,\bx_3)$, $\pi(\bx_2|\bx_1,\bx_3)$, and $\pi(\bx_3|\bx_1,\bx_2)$ that are ``simple'' to sample. Using this decomposition, the GS targeting $\pi$ proceeds as in Algo. \ref{algo:Gibbs}. Somewhat surprisingly, the Markov kernel resulting from concatenating the component-wise kernels admits $\pi(\bx)$ as joint invariant distribution, and thus the chain produced by Algo. \ref{algo:Gibbs} has the desired target density \black{(see \cite[Ch. 10]{casella:robert:2004} for a review of the theory behind this algorithm)}\black. This fundamental property holds even if the simulation from the conditionals is done by using other MCMC algorithms (e.g., RWMH, MALA or HMC steps targeting the conditional densities), though this may result in a deterioration of the algorithm convergence rate. Similarly, the property also holds if the frequency and order of the updates is scheduled randomly and adaptively to improve the overall performance. 

\begin{algorithm}[H]
\caption{Gibbs sampler algorithm}
\begin{algorithmic}
\STATE Set an initial state $\bx^{(0)}=(\bx_1^{(0)},\bx_2^{(0)},\bx_3^{(0)})$
\FOR {$t=1$ to $T$}
\STATE Generate $\bx^{(t)}_1\sim \pi\left(\bx_1 | \bx_2^{(t-1)},\bx_3^{(t-1)}\right)$
\STATE Generate $\bx^{(t)}_2\sim \pi\left(\bx_2 | \bx_1^{(t)}, \bx_3^{(t-1)}\right)$
\STATE Generate $\bx^{(t)}_3\sim \pi\left(\bx_3 | \bx_1^{(t)}, \bx_2^{(t)}\right)$
\ENDFOR
\end{algorithmic}
\label{algo:Gibbs}
\end{algorithm}

As with other MH algorithms, the performance of the GS depends on the correlation structure of $\pi$. Efficient samplers seek to update simultaneously the elements of $\bx$ that are highly correlated with each other, and to update ``slow-moving'' elements more frequently. The structure of $\pi$ can be determined by pilot runs, or alternatively by using an adaptive GS that learns it online and that adapts the updating schedule accordingly as described in \cite{latuszynski2010adaptive}. However, updating elements in parallel often involves simulating from more complicated conditional distributions, and thus introduces a computational overhead.
Finally, it is worth noting that the GS is very useful for SP models, which typically have sparse conditional independence structures (e.g., Markovian properties) and conjugate priors and hyper-priors from the exponential family. This often leads to simple one-dimensional conditional distributions that can be updated in parallel by groups \cite{Bazot_BMC_Bioinfo_2013, Pereyra_IEEE_Trans_IP_2013}. 

\subsection{Partially collapsed Gibbs sampling}
The partially collapsed Gibbs sampler (PCGS) is a recent development in MCMC theory that seeks to address some of the limitations of the conventional GS \cite{dyk:2008a}. As mentioned previously, the GS performs poorly if strongly correlated elements of $\bx$ are updated separately, as this leads to chains that are highly correlated and to an inefficient exploration of the parameter space. However, updating several elements together can also be computationally expensive, particularly if it requires simulating from difficult conditional distributions. In collapsed samplers, this drawback is addressed by carefully replacing one or several conditional densities by \emph{partially collapsed}, or marginalized conditional distributions. 

For illustration, suppose that in our previous example the subvectors $\bx_1$ and $\bx_2$ exhibit strong dependencies, and that as a result the GS of Algo. \ref{algo:Gibbs} performs poorly. Assume that we are able to draw samples from the marginalized conditional density $\pi(\bx_1|\bx_3) = \int \pi(\bx_1,\bx_2|\bx_3) \textrm{d}\bx_2$, which does not depend on $\bx_2$. This leads to the PCGS described in Algo. \ref{algo:PCGibbs} to sample from $\pi$, which ``partially collapses'' Algo. \ref{algo:Gibbs} by replacing $\pi(\bx_1|\bx_2,\bx_3)$ with $\pi(\bx_1|\bx_3)$.
\begin{algorithm}[H]
\caption{Partially collapsed Gibbs sampler}
\begin{algorithmic}
\STATE Set an initial state $\bx^{(0)}=(\bx_1^{(0)},\bx_2^{(0)},\bx_3^{(0)})$
\FOR {$t=1$ to $T$}
\STATE Generate $\bx^{(t)}_1\sim \pi\left(\bx_1 | \bx_3^{(t-1)}\right)$
\STATE Generate $\bx^{(t)}_2\sim \pi\left(\bx_2 | \bx_1^{(t)}, \bx_3^{(t-1)}\right)$
\STATE Generate $\bx^{(t)}_3\sim \pi\left(\bx_3 | \bx_1^{(t)}, \bx_2^{(t)}\right)$
\ENDFOR
\end{algorithmic}
\label{algo:PCGibbs}
\end{algorithm}

Van Dyk and Park \cite{dyk:2008a} established that \black{the PCGS is always at least as efficient as the conventional GS}, \black and it has been observed that \black{that the PCGS is remarkably efficient for some statistical models} \black \cite{dyk:2008b, kail:2012}. Unfortunately, PCGSs are not as widely applicable as GSs because they require simulating exactly from the partially collapsed conditional distributions. In general, using MCMC simulation (e.g., MH steps) within a PCGS will lead to an incorrect MCMC algorithm \cite{dyk:2013}. Similarly, altering the order of the updates (e.g., by permuting the simulations of $\bx_1$ and $\bx_2$ in Algo. \ref{algo:PCGibbs}) may also alter the target density \cite{dyk:2008a}.

\section{Surrogates for stochastic simulation}\label{Sec3}


\newcommand{\setQ}{\mathcal{Q}}
\newcommand{\notj}{_{\setminus j}}
\newcommand{\bethe}{_\textsf{B}}
\newcommand{\neigh}{\mathfrak{N}}
\newcommand{\MF}{_\textsf{MF}}
\newcommand{\secref}[1]{Section~\ref{sec:#1}}
\newcommand{\figref}[1]{Figure~\ref{fig:#1}}
\newcommand{\new}{^\text{\textsf{new}}}
\newcommand{\tmp}{^{\setminus \alpha}}
\newcommand{\cav}{^{\setminus \alpha}}
\newcommand{\msg}[2]{m_{#1\rightarrow #2}}
\newcommand{\site}[2]{m_{#1,#2}}
\newcommand{\pyz}{p_{\textsf{y}|\textsf{z}}}
\newcommand{\px}{p_{\textsf{x}}}
\newcommand{\qxn}{q_{\textsf{x},n}}
\newcommand{\qzm}{q_{\textsf{z},m}}

\let\oldvec\vec
\let\oldeqref\eqref
\let\oldhat\hat
\renewcommand{\vec}[1]{\ensuremath{\boldsymbol{#1}}}
\renewcommand{\eqref}[1]{(\ref{eq:#1})}
\renewcommand{\hat}{\widehat}

\subsection{Variational Bayes} \label{sec:VB}

In the \emph{variational Bayes} (VB) approach described in \cite{Jordan:ML:99,Bishop:Book:07}, the true posterior $p(\vec{x}|\vec{y})$ is approximated by a density $q_\star(\vec{x})\in\setQ$, where $\setQ$ is a subset of valid densities on $\vec{x}$.
In particular, 
\begin{align}
q_\star(\vec{x})
&= \arg\min_{q\in\setQ} \Dkl\big(q(\vec{x})\big\|p(\vec{x}|\vec{y})\big)
\label{eq:VB}
\end{align}
where $\Dkl(q\|p)$ denotes the Kullback-Leibler (KL) divergence between $p$ and $q$.
As a result of the optimization in \eqref{VB} over a function, this is termed ``variational Bayes'' because of the relation to the calculus of variations \cite{Weinstock:Book:74}.
Recalling that $\Dkl(q\|p)$ reaches its minimum value of zero if and only if $p=q$ \cite{Cover:Book:06}, we see that $q_\star(\vec{x})=p(\vec{x}|\vec{y})$ when $\setQ$ includes all valid densities on $\vec{x}$. 
However, the premise is that $p(\vec{x}|\vec{y})$ is too difficult to work with, and so $\setQ$ is chosen as a balance between fidelity and tractability.

Note that the use of $\Dkl(q\|p)$, rather than $\Dkl(p\|q)$, 
implies a search for a $q_\star$ that agrees with the true posterior $p(\vec{x}|\vec{y})$ over the set of $\vec{x}$ where $p(\vec{x}|\vec{y})$ is large.
We will revisit this choice when discussing expectation propagation in \secref{EP}.

Rather than working with the KL divergence directly, it is common to decompose it as follows
\begin{equation}
\Dkl\big(q(\vec{x})\big\|p(\vec{x}|\vec{y})\big)
= \int q(\vec{x}) \ln \frac{q(\vec{x})}{p(\vec{x}|\vec{y})} \dif\vec{x} 
= \ln Z + F(q) \label{eq:Dkl}
\end{equation}
where
\begin{equation}
F(q) \defn \int q(\vec{x}) \ln \frac{q(\vec{x})}{p(\vec{x},\vec{y})} \dif\vec{x}
\label{eq:gibbs}
\end{equation}
is known as the \emph{Gibbs free energy} or \emph{variational free energy}.
Rearranging \eqref{Dkl}, we see that 
\begin{align}
-\ln Z 
= F(q) - \Dkl\big(q(\vec{x})\big\|p(\vec{x}|\vec{y})\big) 
\leq F(q)
\end{align}
as a consequence of $\Dkl\big(q(\vec{x})\big\|p(\vec{x}|\vec{y})\big)\geq 0$.
Thus, $F(q)$ can be interpreted as an upper bound on the negative log partition.
Also, because $\ln Z$ is invariant to $q$, the optimization \eqref{VB} can be rewritten as 
\begin{align}
q_\star(\vec{x})
&= \arg\min_{q\in\setQ} F(q) ,
\label{eq:freemin}
\end{align}
which avoids the difficult integral in \eqref{partition}.
In the sequel, we will discuss several strategies to solve the variational optimization problem \eqref{freemin}.

\subsection{The mean-field approximation} \label{sec:MF}

A common choice of $\setQ$ is the set of fully factorizable densities, resulting in the \emph{mean-field} approximation \cite{Peterson:CS:87,Parisi:Book:88}
\begin{align}
q(\vec{x})
&= \prod_{j=1}^N q_j(x_j) .
\label{eq:MF}
\end{align}
Substituting \eqref{MF} into \eqref{gibbs} yields the mean-field free energy
\begin{equation}\label{eq:MFfree}
\begin{split}
F\MF(q)
\defn& 
\int \Big[\prod_{j} q_j(x_j)\Big] \ln \frac{1}{p(\vec{x},\vec{y})} \dif\vec{x} 
-\sum_{j=1}^N h(q_j)
\end{split}
\end{equation}
where $h(q_j)\defn -\int q_j(x_j) \ln q_j(x_j) \dif x_j$ denotes the differential entropy.
Furthermore, for $j=1,\dots,N$, equation \eqref{MFfree} can be written as
\begin{align}
F\MF(q)
&= \Dkl\big(q_j(x_j) \big\| g_j(x_j,\vec{y}) \big)
-\sum_{i\neq j} h(q_i) 
\label{eq:MFkl} \\
g_j(x_j,\vec{y})
&\defn \exp \int 
\Big[ \prod_{i\neq j} q_i(x_i) \Big] \ln p(\vec{x},\vec{y}) \dif\vec{x}\notj
\label{eq:MFg}
\end{align}
\black{where $\vec{x}\notj \defn [x_1,\dots,x_{j-1},x_j,\dots,x_N]^T$ for $j=1,\dots,N$.
Equation \eqref{MFkl} implies the optimality condition}\black
\begin{align}
q_{j,\star}(x_j)
&= \frac{g_{j,\star}(x_j,\vec{y})}{\int g_{j,\star}(x_j',\vec{y}) \dif x_j'} ~~ \forall j=1,\dots,N
\label{eq:MFopt}
\end{align}
\black{where $q_\star(\vec{x}) = \prod_{j=1}^N q_{j,\star}(x_j)$ and where $g_{j,\star}(x_j,\vec{y})$ is defined as in \eqref{MFg} but with $q_{i,\star}(x_i)$ in place of $q_i(x_i)$.} \black 
Equation \eqref{MFopt} suggests an iterative coordinate-ascent algorithm:
update each component $q_j(x_j)$ of $q(\vec{x})$ while holding the others fixed.
But this requires solving the integral in \eqref{MFg}. 
A solution arises when $\forall j$ the conditionals $p(x_j,\vec{y}|\vec{x}\notj)$ belong to the same exponential family of distributions \cite{Wainwright:FTML:08}, i.e.,
\begin{align}
p(x_j,\vec{y}\,|\,\vec{x}\notj)
&\propto h(x_j) \exp\big(\vec{\eta}(\vec{x}\notj,\vec{y})^T \vec{t}(x_j) \big)
\label{eq:expfam}
\end{align}
where the sufficient statistic $\vec{t}(x_j)$ parameterizes the family.
The exponential family encompasses a broad range of distributions, notably jointly Gaussian and multinomial $p(\vec{x},\vec{y})$.
Plugging $p(\vec{x},\vec{y})=p(x_j,\vec{y}\,|\,\vec{x}\notj) p(\vec{x}\notj,\vec{y})$ and \eqref{expfam} into \eqref{MFg} immediately gives
\begin{align}
g_j(x_j,\vec{y})
&\propto h(x_j) \exp\big(\Ex\{\vec{\eta}(\vec{x}\notj,\vec{y})\}^T \vec{t}(x_j) \big)
\end{align}
\black{where the expectation is taken over $\vec{x}\notj \sim \prod_{i\neq j} q_{i,\star}(x_i)$. Thus,} \black if each $q_j$ is chosen from the same family, i.e., 
$q_j(x_j) \propto h(x_j) \exp\big(\vec{\gamma}_j^T \vec{t}(x_j) \big)~\forall j$,
then \eqref{MFopt} reduces to the moment-matching condition
\begin{align}
\vec{\gamma}_{j,\star}
&= \Ex\big\{\vec{\eta}(\vec{x}\notj,\vec{y})\big\}
\end{align}
\black{where $\vec{\gamma}_{j,\star}$ is the optimal value of $\vec{\gamma}_{j}$}. \black

\subsection{The Bethe approach} \label{sec:Bethe}

In many cases, the fully factored model \eqref{MF} yields too gross of an approximation.
As an alternative, one might try to fit a model $q(\vec{x})$ that has a similar dependency structure as $p(\vec{x}|\vec{y})$. In the sequel, we assume that the true posterior factors as
\begin{align}
p(\vec{x}|\vec{y})
&= Z^{-1} \prod_\alpha f_\alpha(\vec{x}_\alpha)
\label{eq:factor}
\end{align}
where $\vec{x}_\alpha$ are subvectors of $\vec{x}$ (sometimes called \emph{cliques} or \emph{outer clusters}) and $f_\alpha$ are non-negative potential functions.
\black{Note that the factorization} \ModifPS{\eqref{factor} defines a Gibbs random field when $p(\vec{x}|\vec{y})>0$.} \black
When a collection of variables $\{x_n\}$ always appears together in a factor, we can collect them into $\vec{x}_\beta$, an \emph{inner cluster}, although it is not necessary to do so. 
For simplicity we will assume that these $\vec{x}_\beta$ are non-overlapping (i.e., $\vec{x}_\beta\cap \vec{x}_{\beta'}=0~\forall \beta\neq\beta'$), so that $\{\vec{x}_\beta\}$ represents a partition of $\vec{x}$.
The factorization \eqref{factor} can then be drawn as a \emph{factor graph} to help visualize the structure of the posterior, as in \figref{factor_graph}.

\begin{figure}[ht!]
\centering
\psfragfig[width=2.0in]{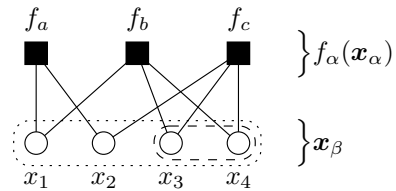}
{
\psfrag{f1}[][Bl][1.0]{$f_a$}
\psfrag{f2}[][Bl][1.0]{$f_b$}
\psfrag{f3}[][Bl][1.0]{$f_c$}
\psfrag{fx1}[b][Bl][1.0]{$x_1$}
\psfrag{fx2}[b][Bl][1.0]{$x_2$}
\psfrag{fx3}[b][Bl][1.0]{$x_3$}
\psfrag{fx4}[b][Bl][1.0]{$x_4$}
\psfrag{fxalpha}[l][Bl][1.0]{$\Big\} f_\alpha(\boldsymbol{x}_\alpha)$~}
\psfrag{fxbeta}[l][Bl][1.0]{$\Big\} \boldsymbol{x}_\beta$~}
}
\caption{
An example of a factor graph, which is a bipartite graph consisting of variable nodes, (circles/ovals), and factor nodes, (boxes).
In this example, $\vec{x}_{a}=\{x_1,x_2\}$,
$\vec{x}_{b}=\{x_1,x_3,x_4\}$,
$\vec{x}_{c}=\{x_2,x_3,x_4\}$.
There are several choices for the inner clusters $\vec{x}_\beta$.
One is the full factorization  
$\vec{x}_{1}=x_1$, 
$\vec{x}_{2}=x_2$, 
$\vec{x}_{3}=x_3$, and 
$\vec{x}_{4}=x_4$.
Another is the partial factorization
$\vec{x}_{1}=x_1$, 
$\vec{x}_{2}=x_2$, and
$\vec{x}_{3}=\{x_3,x_4\}$,
which results in the ``super node'' in the dashed oval.
Another is no factorization:
$\vec{x}_{1}=\{x_1,x_2,x_3,x_4\}$, 
resulting in the ``super node'' in the dotted oval.
In the latter case, we redefine each factor $f_\alpha$ to have the full domain $\vec{x}$ (with trivial dependencies where needed).
}
\label{fig:factor_graph}
\end{figure}

We now seek a tractable way to build an approximation $q(\vec{x})$ with the same dependency structure as \eqref{factor}.
But rather than designing $q(\vec{x})$ as a whole, we design the cluster marginals, $\{q_\alpha(\vec{x}_\alpha)\}$ and $\{q_\beta(\vec{x}_\beta)\}$, which must be non-negative, normalized, and consistent
\begin{align} 
  0 &\leq q_\alpha(\vec{x}_\alpha),
  ~~ 0 \leq q_\beta(\vec{x}_\beta) ~~\forall \alpha,\beta,\vec{x}_\alpha,\vec{x}_\beta
        \label{eq:positivity}\\
  1 &= \int q_\alpha(\vec{x}_\alpha) \dif \vec{x}_\alpha 
  ~~ 1 = \int q_\beta(\vec{x}_\beta) \dif \vec{x}_\beta ~~\forall \alpha,\beta
        \label{eq:normalization}\\
    & q_\beta(\vec{x}_\beta) 
  = \int q_\alpha(\vec{x}_\alpha) \dif \vec{x}_{\alpha \setminus \beta} 
  ~~\forall \alpha,\beta\in\neigh_\alpha,\vec{x}_\beta
        \label{eq:consistency}
\end{align} 
where \black{$\vec{x}_{\alpha \setminus \beta}$ gathers the components of $\vec{x}$ that are contained in the cluster $\alpha$ and not in the cluster $\beta$}, \black and $\neigh_\alpha$ denotes the neighborhood of the factor $\alpha$ (\ModifPS{i.e., the set of inner clusters $\beta$ connected to $\alpha$}).

In general, it is difficult to specify $q(\vec{x})$ from its cluster marginals.
However, in the special case that the factor graph has a tree structure (i.e., there is at most one path from one node in the graph to another), we have \cite{Yedidia:TIT:05} 
\begin{align} 
q(\vec{x}) 
&= \frac{ \prod_\alpha q_\alpha(\vec{x}_\alpha) }
   { \prod_{\beta} q_\beta(\vec{x}_\beta)^{N_\beta-1} }
\label{eq:tree}
\end{align} 
where $N_\beta=|\neigh_\beta|$ is the neighborhood size of the cluster $\beta$.
In this case, the free energy \eqref{gibbs} simplifies to
\begin{align} 
F(q)
&= \underbrace{
  \sum_\alpha \Dkl\big(q_\alpha \big\| f_\alpha\big) + \sum_\beta (N_\beta-1) h(q_\beta) 
}_{\displaystyle \defn F\bethe\big(\{q_\alpha\},\{q_\beta\}\big)}
+ \text{const},
\label{eq:BFE}
\end{align} 
where $F\bethe$ is known as the \emph{Bethe free energy} (BFE) \cite{Yedidia:TIT:05}.

Clearly, if the true posterior $\{f_\alpha\}$ has a tree structure, and no constraints beyond \eqref{positivity}-\eqref{consistency} are placed on the cluster marginals $\{q_\alpha\},\{q_\beta\}$, then minimization of $F\bethe\big(\{q_\alpha\},\{q_\beta\}\big)$ will recover the cluster marginals of the true posterior.
But even when $\{f_\alpha\}$ is not a tree, 
$F\bethe\big(\{q_\alpha\},\{q_\beta\}\big)$ can be used as an \emph{approximation} of the Gibbs free energy $F(q)$, and minimizing $F\bethe$ can be interpreted as designing a $q$ that \emph{locally} matches the true posterior.

The remaining question is how to efficiently minimize $F\bethe\big(\{q_\alpha\},\{q_\beta\}\big)$ subject to the (linear) constraints \eqref{positivity}-\eqref{consistency}.
Complicating matters is the fact that $F\bethe\big(\{q_\alpha\},\{q_\beta\}\big)$ is the sum of convex KL divergences and concave entropies.
One option is direct minimization using a ``double loop'' approach like the concave-convex procedure (CCCP) \cite{Yuille:NIPS:02}, where the outer loop linearizes the concave term about the current estimate and the inner loop solves the resulting convex optimization problem (typically with an iterative technique).
Another option is belief propagation, which is described below.

\subsection{Belief propagation} \label{sec:BP}

\emph{Belief propagation} (BP) \cite{Gallager:TIT:62,Pearl:Book:88} is an algorithm for computing (or approximating) marginal probability density functions (pdfs)\footnote{Note that another form of BP exists to compute the maximum a posteriori (MAP) estimate $\arg\max_{\vec{x}} p(\vec{x}|\vec{y})$ known as the ``max-product'' or ``min-sum'' algorithm \cite{Pearl:Book:88}. However, this approach does not address the problem of computing surrogates for stochastic methods, and so is not discussed further.} 
like $q_\beta(\vec{x}_\beta)$ and $q_\alpha(\vec{x}_\alpha)$ by propagating messages on a factor graph.
The standard form of BP is given by the \emph{sum-product algorithm} (SPA) \cite{Kschischang:TIT:01}, which computes the following messages from each factor node $f_\alpha$ to each variable (super) node $\vec{x}_\beta$ and vice versa
\begin{align}
\msg{\alpha}{\beta}(\vec{x}_\beta)
&\leftarrow \int f_\alpha(\vec{x}_\alpha) 
        \prod_{\beta'\in{\neigh_\alpha}\setminus\beta} 
        \msg{\beta'}{\alpha}(\vec{x}_{\beta'}) 
        \dif \vec{x}_{\alpha\setminus\beta} 
        \label{eq:msgab}\\
\msg{\beta}{\alpha}(\vec{x}_\beta)
&\leftarrow \prod_{\alpha'\in\neigh_\beta\setminus\alpha} 
        \msg{\alpha'}{\beta}(\vec{x}_\beta) .
        \label{eq:msgba}
\end{align}
These messages are then used to compute the \emph{beliefs} 
\begin{align}
q_\beta(\vec{x}_\beta)
&\propto \prod_{\alpha\in\neigh_\beta} \msg{\alpha}{\beta}(\vec{x}_\beta) 
        \label{eq:beliefb}\\
q_\alpha(\vec{x}_\alpha)
&\propto f_\alpha(\vec{x}_\alpha) 
        \prod_{\beta\in\neigh_\alpha} 
        \msg{\beta}{\alpha}(\vec{x}_\beta)
        \label{eq:beliefa}
\end{align}
which must be normalized in accordance with \eqref{normalization}.
The messages \eqref{msgab}-\eqref{msgba} do not need to be normalized, although it is often done in practice to prevent numerical overflow.

When the factor graph $\{f_\alpha\}$ has a tree structure, the BP-computed marginals coincide with the true marginals after one round of forward and backward message passes. Thus, BP on a tree-graph is sometimes referred to as the \emph{forward-backward} algorithm, particularly in the context of hidden Markov models \cite{Rabiner:PROC:89}.
In the tree case, BP is akin to a dynamic programming algorithm that organizes the computations needed for marginal evaluation in a tractable manner.

When the factor graph $\{f_\alpha\}$ has cycles or ``loops,'' BP can still be applied by iterating the message computations \eqref{msgab}-\eqref{msgba} until convergence (not guaranteed), which is known as \emph{loopy BP} (LBP).
However, the corresponding beliefs \eqref{beliefb}-\eqref{beliefa} \black{are in general only approximations} \black of the true marginals.
This suboptimality is expected because exact marginal computation on a loopy graph is an NP-hard problem \cite{Cooper:AI:90}. 
Still, the answers computed by LBP are in many cases very accurate \cite{Murphy:UAI:99}.
For example,
LBP methods have been successfully applied to communication and SP problems such as:
turbo decoding \cite{McEliece:JSAC:98},
LDPC decoding \cite{Gallager:TIT:62,MacKay:Book:03},
inference on Markov random fields \cite{Freeman:IJCV:00},
multiuser detection \cite{Boutros:TIT:02}, 
and compressive sensing \cite{Donoho:ITW:10a,Rangan:ISIT:11}.

Although the excellent performance of LBP was at first a mystery, it was later established that LBP minimizes the constrained BFE.
More precisely, the fixed points of LBP coincide with the stationary points of $F\bethe\big(\{q_\alpha\},\{q_\beta\}\big)$ from \eqref{BFE} under the constraints \eqref{positivity}-\eqref{consistency} \cite{Yedidia:TIT:05}.
The link between LBP and BFE can be established through the Lagrangian formalism, which converts constrained BFE minimization to an unconstrained minimization through the incorporation of additional variables known as Lagrange multipliers  \cite{Bertsekas:Book:99}.
By setting the derivatives of the Lagrangian to zero, one obtains a set of equations that are equivalent to the message updates \eqref{msgab}-\eqref{msgba} \cite{Yedidia:TIT:05}. 
In particular, the stationary-point versions of the Lagrange multipliers equal the fixed-point versions of the loopy SPA log-messages.

Note that, unlike the mean-field approach \eqref{MF}, the cluster-based nature of LBP does not facilitate an explicit description of the joint-posterior approximation $q\in\setQ$ from \eqref{VB}. 
The reason is that, when the factor graph is loopy, there is no straightforward relationship between the joint posterior $q(\vec{x})$ and the cluster marginals $\{q_\alpha(\vec{x}_\alpha)\},\{q_\beta(\vec{x}_\beta)\}$, as explained before \eqref{tree}. 
Instead, it is better to interpret LBP as an efficient implementation of the Bethe approach from \secref{Bethe}, which aims for a local approximation of the true posterior.

In summary, by constructing a factor graph with low-dimensional $\{\vec{x}_\beta\}$ and applying BP or LBP, we trade the high-dimensional integral  $q_\beta(\vec{x}_\beta)=\int p(\vec{x}|\vec{y}) \dif \vec{x}_{\setminus \beta}$ for a sequence of low-dimensional message computations \eqref{msgab}-\eqref{msgba}.  
But \eqref{msgab}-\eqref{msgba} are themselves tractable only for a few families of $\{f_\alpha\}$.
Typically, $\{f_\alpha\}$ are limited to members of the exponential family closed under marginalization (see \cite{Heskes:JSM:05}), so that the updates of the message pdfs \eqref{msgab}-\eqref{msgba} reduce to updates of the natural parameters (i.e., $\vec{\eta}$ in \eqref{expfam}).
The two most common instances are multivariate Gaussian pdfs and multinomial probability mass functions (pmfs).
For both of these cases, when LBP converges, it tends to be much faster than double-loop algorithms like CCCP (see, e.g., \cite{Heskes:NIPS:04}).  
However, LBP does not always converge \cite{Murphy:UAI:99}.

\subsection{Expectation propagation} \label{sec:EP}

\emph{Expectation propagation} (EP) \cite{Minka:Diss:01} (see also the overviews \cite{Heskes:JSM:05,Seeger:Tech:05}) is an iterative method of approximate inference that is reminiscent of LBP but has much more flexibility with regards to the modeling distributions. In EP, the true posterior $p(\vec{x}|\vec{y})$, which is assumed to factorize as in \eqref{factor}, is approximated by $q(\vec{x})$ such that 
\begin{align}
q(\vec{x}) 
&\propto \prod_{\alpha} m_\alpha(\vec{x}_\alpha) 
\label{eq:EPalpha}
\end{align}
where $\vec{x}_\alpha$ are the same as in \eqref{factor} and $m_\alpha$ are referred to as ``site approximations.''
Although no constraints are imposed on the true-posterior factors $\{f_\alpha\}$, the approximation $q(\vec{x})$ is restricted to a factorized exponential family.
In particular,
\begin{align}
q(\vec{x}) 
&= \prod_{\beta} q_\beta(\vec{x}_\beta) 
\label{eq:EPbeta}\\
q_\beta(\vec{x}_\beta) 
&= \exp\big(\vec{\gamma}_\beta^T \vec{t}_\beta(\vec{x}_\beta) - c_\beta(\vec{\gamma}_\beta)\big), ~\forall \beta
\label{eq:EPexpfam} ,
\end{align}
with some given base measure.
We note that our description of EP applies to arbitrary partitions $\{\vec{x}_\beta\}$, from the trivial partition $\vec{x}_\beta=\vec{x}$ to the full partition $\vec{x}_\beta=x_\beta$.

The EP algorithm iterates the following updates over all factors $\alpha$ until convergence (not guaranteed)
\begin{align}
q\cav(\vec{x})
&\leftarrow q(\vec{x})/m_\alpha(\vec{x}_\alpha)
\label{eq:qcav}\\
\hat{q}\tmp(\vec{x})
&\leftarrow q\cav(\vec{x}) f_\alpha(\vec{x}_\alpha)
\label{eq:qtmp}\\
q\new(\vec{x})
&\leftarrow \arg\min_{q\in \setQ} \Dkl\big(\hat{q}\tmp(\vec{x})\big\|q(\vec{x})\big) 
\label{eq:qnew}\\
m_\alpha\new(\vec{x}_\alpha)
&\leftarrow q\new(\vec{x})/q\cav(\vec{x}) 
\label{eq:mnew}\\
q(\vec{x})
&\leftarrow q\new(\vec{x})  
\label{eq:EPq} \\
m_\alpha(\vec{x}_\alpha)
&\leftarrow m_\alpha\new(\vec{x}_\alpha) 
\label{eq:EPm}
\end{align}
where $\setQ$ in \eqref{qnew} refers to the set of $q(\vec{x})$ obeying \eqref{EPbeta}-\eqref{EPexpfam}.
Essentially, \eqref{qcav} removes the $\alpha$th site approximation $m_\alpha$ from the posterior model $q$, and \eqref{qtmp} replaces it with the true factor $f_\alpha$. 
Here, $q\cav$ is know as the ``cavity'' distribution.
The quantity $\hat{q}\tmp$ is then projected onto the exponential family in \eqref{qnew}.
The site approximation is then updated in \eqref{mnew}, and the old quantities are overwritten in \eqref{EPq}-\eqref{EPm}.
Note that the right side of \eqref{mnew} depends only on $\vec{x}_\alpha$ because $q\new(\vec{x})$ and $q\cav(\vec{x})$ differ only over $\vec{x}_\alpha$.
Note also that the KL divergence in \eqref{qnew} is reversed relative to \eqref{VB}.

The EP updates \eqref{qtmp}-\eqref{EPm} can be simplified by leveraging the factorized exponential family structure in \eqref{EPbeta}-\eqref{EPexpfam}. 
First, for \eqref{EPalpha} to be consistent with \eqref{EPbeta}-\eqref{EPexpfam}, each site approximation must factor into exponential-family terms, i.e.,
\begin{align}
m_\alpha(\vec{x}_\alpha)
&= \prod_{\beta\in\neigh_\alpha} \site{\alpha}{\beta}(\vec{x}_\alpha) \\
\site{\alpha}{\beta}(\vec{x}_\alpha)
&= \exp\big(\vec{\mu}_{\alpha,\beta}^T\vec{t}_\beta(\vec{x}_\beta)\big) .
\end{align}
It can then be shown \cite{Heskes:JSM:05} that \eqref{qcav}-\eqref{qnew} reduce to 
\begin{align}
\vec{\gamma}_\beta\new 
&\leftarrow \arg\max_{\vec{\gamma}} \left[ \vec{\gamma}^T \Ex_{\hat{q}\tmp}\{\vec{t}_\beta(\vec{x}_\beta)\} - c_\beta(\vec{\gamma})\right]
\end{align}
for all $\beta\in\neigh_\alpha$,
which can be interpreted as the moment matching condition 
$\Ex_{q\new}\{\vec{t}_\beta(\vec{x}_\beta)\}
= \Ex_{\hat{q}\tmp}\{\vec{t}_\beta(\vec{x}_\beta)\}$.
Furthermore, \eqref{mnew} reduces to
\begin{align}
\vec{\mu}_{\alpha,\beta}\new
&\leftarrow \vec{\gamma}_\beta\new - \sum_{\alpha'\in\neigh_\beta\setminus\alpha}
        \vec{\mu}_{\alpha',\beta}  
\end{align}
for all $\beta\in\neigh_\alpha$.
Finally, \eqref{EPq} and \eqref{EPm} reduce to 
$\vec{\gamma}_\beta \leftarrow \vec{\gamma}_\beta\new$
and
$\vec{\mu}_{\alpha,\beta} \leftarrow \vec{\mu}_{\alpha,\beta}\new$,
respectively, for all $\beta\in\neigh_\alpha$.


Interestingly, in the case that the true factors $\{f_\alpha\}$ are members of an exponential family closed under marginalization, the version of EP described above is equivalent to the SPA up to a change in message schedule.
In particular, for each given factor node $\alpha$, the SPA updates the outgoing message towards one variable node $\beta$ per iteration, whereas EP simultaneously updates the outgoing messages in all directions, resulting in $m\new_\alpha$ (see, e.g., \cite{Bishop:Book:07}).
By restricting the optimization in \eqref{mnew} to a single factor $q\new_\beta$, EP can be made equivalent to the SPA.
On the other hand, for generic factors $\{f_\alpha\}$, EP can be viewed as a tractable approximation of the (intractable) SPA.

Although the above form of EP iterates serially through the factor nodes $\alpha$, it is also possible to perform the updates in parallel, resulting in what is known as the \emph{expectation consistent} (EC) approximation algorithm \cite{Opper:NIPS:05}.

EP and EC have an interesting BFE interpretation.
Whereas the fixed points of LBP coincide with the stationary points of the BFE \eqref{BFE} subject to \eqref{positivity}-\eqref{normalization} and \emph{strong} consistency \eqref{consistency}, the fixed points of EP and EC coincide with the stationary points of the BFE \eqref{BFE} subject to \eqref{positivity}-\eqref{normalization} and the \emph{weak} consistency (i.e., moment-matching) constraint \cite{Heskes:UAI:02}
\begin{align}
\Ex_{\hat{q}\tmp}\{\vec{t}(\vec{x}_\beta)\}
&= \Ex_{q_\beta}\{\vec{t}(\vec{x}_\beta)\},
~\forall \alpha, \beta\in\neigh_\alpha .
\end{align}
EP, like LBP, is not guaranteed to converge.  
Hence, provably convergence double-loop algorithms have been proposed that directly minimize the weakly constrained BFE, e.g., \cite{Heskes:UAI:02}.

\subsection{Approximate message passing} \label{sec:AMP}

So-called \emph{approximate message passing} (AMP) algorithms \cite{Donoho:ITW:10a,Rangan:ISIT:11} have recently been developed for the separable generalized linear model
\begin{align}
p(\vec{x}) = \prod_{j=1}^N \px(x_j) 
&,~~
p(\vec{y}|\vec{x}) = \prod_{m=1}^M \pyz(y_m|\vec{a}^T_m\vec{x})  
\label{eq:GLM}
\end{align}
where the prior $p(\vec{x})$ \black is fully factorizable, as is the conditional pdf $p(\vec{y}|\vec{z})$ relating the observation vector $\vec{y}$ to the (hidden) transform output vector $\vec{z}\defn\vec{A}\vec{x}$, where $\vec{A}\defn[\vec{a}_1,...,\vec{a}_M]^T\in\Real^{M\times N}$ is a known linear transform.
Like EP, AMP allows tractable inference under \emph{generic}\footnote{More precisely, the AMP algorithm \cite{Donoho:ITW:10a} handles Gaussian $\pyz$ while the generalized AMP (GAMP) algorithm \cite{Rangan:ISIT:11} handles arbitrary $\pyz$.} 
$\px$ and $\pyz$.

AMP can be derived as an approximation of LBP on the factor graph constructed with inner clusters $\vec{x}_\beta = x_\beta$ for $\beta=1,\dots,N$, \ModifPS{with outer clusters $\vec{x}_\alpha=\vec{x}$ for $\alpha=1,\dots,M$ and $\vec{x}_\alpha=x_{\alpha-M}$ for $\alpha=M+1,\dots,M+N$,} and \ModifPS{with factors}
\begin{align}
f_\alpha(\vec{x}_\alpha) &= 
\begin{cases}
\pyz(y_\alpha|\vec{a}_\alpha^T\vec{x}) & \alpha=1,\dots,M \\
\px(x_{\alpha-\ModifPS{M}}) & \alpha=\ModifPS{M+1,\dots,M+N} .
\end{cases}
\end{align}
In the large-system limit (LSL), i.e., $M,N\rightarrow\infty$ for fixed ratio $M/N$, the LBP \ModifPS{beliefs $q_\beta(x_\beta)$ simplify to 
\begin{align} 
 q_\beta(x_\beta) &\propto \px(x_\beta) \Normpdf(x_\beta;\hat{r}_\beta,\tau) 
 \label{eq:AMPpx} 
\end{align} 
where $\{\hat{r}_\beta\}_{\beta=1}^N$ and $\tau$ are iteratively updated parameters.
Similarly, for $m=1,\dots,M$, the belief on $z_m$, denoted by $\qzm(\cdot)$, simplifies to
\begin{align} 
 \qzm(z_m) &\propto \pyz(y_m|z_m) \Normpdf(z_m;\hat{p}_m,\nu) 
 \label{eq:AMPpz}  
\end{align} 
where $\{\hat{p}_m\}_{m=1}^M$ and $\nu$ are iteratively updated parameters.}
Each AMP iteration requires only one evaluation of the mean and variance of \eqref{AMPpx}-\eqref{AMPpz}, 
one multiplication by $\vec{A}$ and $\vec{A}^T$, and relatively few iterations, making it very computationally efficient, especially when these multiplications have fast implementations (e.g., \black{using fast Fourier transforms and discrete wavelet transforms} \black).

In the LSL under i.i.d sub-Gaussian $\vec{A}$, AMP is fully characterized by a scalar state evolution (SE).  
When this SE has a unique fixed point, the \ModifPS{marginal} posterior approximations \eqref{AMPpx}-\eqref{AMPpz} are known to be exact \cite{Bayati:TIT:11,Javanmard:II:13}. 

For generic $\vec{A}$, AMP's fixed points coincide with the stationary points of an LSL version of the BFE \cite{Rangan:ISIT:13,Krzakala:ISIT:14}.
When AMP converges, its posterior approximations are often very accurate (e.g., \cite{Vila:TSP:14}), but AMP does not always converge.
In the special case of Gaussian likelihood $\pyz$ and prior $\px$, AMP convergence is fully understood: convergence depends on the ratio of peak-to-average squared singular values of $\vec{A}$, and convergence can be guaranteed for any $\vec{A}$ with appropriate damping \cite{Rangan:ISIT:14}.
For generic $\px$ and $\pyz$, damping greatly helps convergence \cite{Vila:ICASSP:15} but theoretical guarantees are lacking.
A double-loop algorithm to directly minimize the LSL-BFE was recently proposed and shown to have global convergence for strictly log-concave $\px$ and $\pyz$ under generic $\vec{A}$ \cite{Rangan:15}.

\let\vec\oldvec
\let\eqref\oldeqref
\let\hat\oldhat

\section{Optimization methods}\label{Sec4}

\subsection{Optimization problem}
The Monte Carlo methods described in Section \ref{Sec2} provide a general approach for estimating reliably posterior probabilities and expectations. However, their high computational cost often makes them unattractive for applications involving very high dimensionality or tight computing time constraints. One alternative strategy is to perform inference approximately by using deterministic surrogates as described in Section \ref{Sec3}. Unfortunately, these faster inference methods are not as generally applicable, and because they rely on approximations, the resulting inferences can suffer from estimation bias. 
\black{As already mentioned, if one focuses on the MAP estimator, efficient optimization techniques can be employed, which are often more computationally tractable than 
MCMC methods and, for which strong guarantees of convergence exist.}
In many SP applications, the computation of the MAP estimator of $\bx$ can be formulated as an optimization problem having the following form
\begin{equation}\label{e:optMAp}
\minimize{\bx\in \RR^N}{\varphi(\bH\bx,\by) + g(\bD\bx)}
\end{equation}
where $\varphi\colon \RR^M\times \RR^M \to \RX$, $g\colon \RR^P \to \RX$, $\bH\in \RR^{M\times N}$,
and $\bD\in \RR^{P \times N}$ \black{with $P\in \mathbb{N}^*$}.
For example, $\bH$ may be a linear operator modeling
a degradation of the signal of interest, $\by$ a vector of observed data, $\varphi$ a least-squares criterion corresponding
to the negative log-likelihood \black{associated with} \black an additive zero-mean white Gaussian noise, $g$ a sparsity promoting measure, e.g., an
$\ell_1$ norm, and $\bD$ a frame analysis transform or a gradient operator.

Often, $\varphi$ is an additively separable function, i.e.,
\begin{equation} \label{e:seph}
\varphi(\bz,\by) = \frac{1}{M} \sum_{i = 1} ^M \varphi_i(z_i,y_i)
\quad\forall (\bz,\by) \in (\RR^M)^2\quad
\end{equation}
where $\boldsymbol{z}= [z_1,...,z_M]^T$.
Under this condition, the previous optimization problem becomes an instance of the more general stochastic one
\begin{equation}\label{e:probminstochgen}
\minimize{\bx \in \RR^N}{\Phi(\bx) +g(\bD\bx)}
\end{equation}
involving the expectation
\begin{equation}\label{e:expectedterm}
\Phi(\bx) = \E\{\varphi_j(\bh_j^T \bx,y_j)\}
\end{equation}
where $j$, $\by$, and $\bH$ are now assumed to be random variables \black{and the expectation is computed with respect to the joint distribution of $(j,\by,\bH)$,} \black with
$\bh_j^T$ the  $j$-th line of $\bH$. 
 More precisely, when \eqref{e:seph} holds,
\eqref{e:optMAp} is then a special case of \eqref{e:probminstochgen} with $j$ uniformly distributed over $\{1,\ldots,M\}$
and $(\by,\bH)$ deterministic. Conversely, it is also possible to consider that $j$ is deterministic and that for every $i\in \{2,\ldots,M\}$, $\varphi_i = \varphi_1$, and
$(y_i,\bh_i)_{1\le i \le M}$ are identically distributed random variables.
In this second scenario, because of  the separability condition \eqref{e:seph}, \black{the optimization problem}~\eqref{e:optMAp}  \black can be regarded as a proxy for \eqref{e:probminstochgen}, where the expectation $\Phi(\bx)$ is approximated by a sample estimate (or stochastic approximation under suitable mixing assumptions). All these remarks illustrate the  existing connections between problems \eqref{e:optMAp} and \eqref{e:probminstochgen}.\\
Note that the stochastic optimization problem defined in \eqref{e:probminstochgen} has been extensively investigated in two communities: machine learning,
and adaptive filtering, often under quite different practical assumptions on the forms of the functions $(\varphi_j)_{j\ge 1}$ and $g$.
In machine learning \cite{Bottou2004,ML2011,Theodoridis_S_2015_Machine_L}, $\bx$ indeed represents the vector of parameters of a classifier which has to be learnt, whereas in adaptive filtering 
\cite{Haykin2002,Sayed_A-H_2011_book_Adaptive_f}, it is generally the impulse response of
an unknown filter which needs to be identified and possibly tracked. In order to simplify our presentation, in the rest of this section, we will
assume that the functions $(\varphi_j)_{j \ge 1}$ are convex and Lipschitz-differentiable with respect to their first argument (for example, they may be logistic functions).

\subsection{Optimization algorithms for solving stochastic problems}

The main difficulty arising in the resolution of the stochastic optimization problem \eqref{e:probminstochgen} is that the integral involved in the expectation term often cannot be computed in practice since it is generally high-dimensional and the underlying probability measure is usually unknown. Two main computational approaches have been proposed in the literature to overcome this issue. The first idea
is to approximate the expected loss function by using a finite set of observations and to minimize the associated empirical loss \eqref{e:optMAp}. The resulting deterministic optimization problem can then be solved by using either deterministic or stochastic algorithms, the latter being the topic of Section~\ref{se:stochalgpbdet}. Here, we focus on the second family of methods grounded in stochastic approximation techniques to handle the expectation in \eqref{e:expectedterm}. More precisely, a sequence of identically distributed samples $(y_j,\bh_j)_{j \ge 1}$ is drawn, which are processed sequentially 
according to some update rule. The iterative algorithm  aims to produce a sequence of random iterates $(\bx_j)_{j \ge 1}$ converging 
to a solution to \eqref{e:probminstochgen}.

We begin with a group of \emph{online} learning algorithms based on extensions of the well-known \emph{stochastic gradient descent} (SGD) approach. Then we will turn our attention to stochastic optimization techniques developed in the context of adaptive filtering.
 
\subsubsection{Online learning methods based on SGD}

Let us assume that 
an estimate $\bu_j \in \RR^N$ of the gradient of $\Phi$ at $\bx_j$ is available at each iteration $j \ge 1$.
A popular strategy for solving \eqref{e:probminstochgen} in this context leverages the gradient estimates to derive a so-called \emph{stochastic forward-backward} (SFB) scheme, (also sometimes called \emph{stochastic proximal gradient algorithm})
\begin{align}
(\forall j \ge 1) & \quad \bz_j = \prox_{\gamma_j g \circ \bD} \left(\bx_j - \gamma_j \bu_j \right)\nonumber\\ & \quad \bx_{j+1} = (1 - \lambda_j) \bx_j + \lambda_j \bz_j
\label{es:StochFB}
\end{align}
where $(\gamma_j)_{j \ge 1}$ is a sequence of positive stepsize values and $(\lambda_j)_{j \ge 1}$ is a sequence of relaxation parameters in $]0,1]$. Hereabove, $\prox_{\black{\psi}}(\bv)$ denotes the proximity operator at $\bv\in \RR^N$ of a lower-semicontinuous convex function 
$\psi\colon\RR^N \to ]-\infty,+\infty]$ with nonempty domain, \black{i.e.,} \black the unique minimizer of\linebreak $\psi + \frac{1}{2} \| \cdot - \bv \|^2$ 
(see 
\cite{Combettes_2010} and the references therein), \black{and $g \circ \bD(\bx)= g(\bD \bx)$}\black. 
A convergence analysis of \black{the SFB scheme} \black has been conducted in 
\cite{Duchi_J_2009_jmlr_efficient_oblufbs,Atchade_Y_2014_Stochastic_pgga,Rosasco_L_2014_Stochastic_fbsmsmiih,Combettes_P_2014_Stochastic}, under various assumptions on the functions $\Phi$, $g$, on the stepsize sequence, and on the statistical properties of $(\bu_j)_{j \ge 1}$. 
For example, if $\bx_1$ is set to a given (deterministic) value, the sequence $(\bx_j)_{j \ge 1}$ generated by \eqref{es:StochFB} is guaranteed to converge almost surely to a solution of Problem \eqref{e:probminstochgen} under the following technical assumptions \cite{Rosasco_L_2014_Stochastic_fbsmsmiih}
\begin{itemize}
 \item[(i)] $\Phi$ has a $\beta^{-1}$-Lipschitzian gradient with $\beta \in ]0,+\infty[$, 
  $g$ is a lower-semicontinuous convex function, and $\Phi + g \circ \bD$ is strongly convex.
	\item[(ii)] For every $j \ge 1$, 
	\begin{align*}
	&  \Ex{\{\|\bu_j\|^2\}} < +\infty, \;\ModifJCP{\Ex\{{\bu_j}\mid {\XX_{j-1}}\} = \nabla \Phi (\bx_j)},\\
& \ModifJCP{\Ex\{{\|\bu_j -  \nabla \Phi (\bx_j)\|^2} \mid{\XX_{j-1}}\}} \leq \sigma^2 (1 + \alpha_j \|  \nabla \Phi (\bx_j)\|^2)
\end{align*}
{\ModifJCP{where $\XX_{j} = (y_i,\bh_i)_{1\le i \le j}$}}, and
$\alpha_j$ and $\sigma$ are positive values such that $\gamma_j \leq (2 - \epsilon) \beta (1 + 2 \sigma^2 \alpha_j)^{-1}$ with $\epsilon > 0$.
	\item[(iii)] 
	We have
	$$ 
	\sum_{j \ge 1} \lambda_j \gamma_j = + \infty \quad \text{and} \quad \sum_{j \ge 1} \chi_j^2 < + \infty
	$$
	where, for every $j \ge 1$, $\chi_j^2 = \lambda_j \gamma_j^2 (1 + 2  \alpha_j \| \nabla \Phi (\overline{\bx})\|^2)$
	and $\overline{\bx}$ is the solution of Problem~\eqref{e:probminstochgen}.
\end{itemize}

%
%
%
 
When $g \equiv 0$, the SFB algorithm in \eqref{es:StochFB} becomes equivalent to SGD 
\cite{Robbins_H_1951_Stochastic_am,Ermo67,Guse71,Bertsekas_D_20000-siopt_Gradient_cgme}. According to the above result, the convergence of SGD is ensured as soon as $\sum_{j \ge 1} \lambda_j \gamma_j = + \infty$ and $\sum_{j \ge 1} \lambda_j \gamma_j^2 < + \infty$. In the unrelaxed case defined by $\lambda_j \equiv 1$, we then retrieve a particular case of the decaying condition $\gamma_j \propto j^{-1/2-\delta}$ with $\delta \in ]0,1/2]$ usually  imposed on the stepsize in the convergence studies of SGD under slightly different assumptions on the gradient estimates $(\bu_j)_{j \ge 1}$ (see \cite{Bach2011,Nemirovski08} for more details).  Note also that better convergence properties can be obtained, if a Polyak-Ruppert averaging approach is performed, \black{i.e.,} the averaged sequence $\left(\overline{\bx}_j\right)_{j \ge 1}$, defined \black{as $\overline{\bx}_j= \frac{1}{j} \sum_{i = 1}^j \bx_i$ for every $j \ge 1$}, \black is considered instead of $(\bx_j)_{j \ge 1}$ in the convergence analysis \cite{Polyak92,Bach2011}.  

We now comment on approaches related to SFB that have been proposed in the literature to solve \eqref{e:probminstochgen}. It should first be noted that a simple alternative strategy to deal with a possibly nonsmooth term $g$ is to incorporate a subgradient step into the previously mentioned SGD algorithm \cite{Shalev07}. However, this approach, like its deterministic version, may suffer from a slow convergence rate \cite{Xiao10}. Another family of methods, close to SFB, adopt the \emph{regularized dual averaging} (RDA) strategy, first introduced in \cite{Xiao10}. The principal difference between SFB and RDA methods is that the latter rely on \ModifJCP{iterative} averaging of the stochastic gradient estimates, which consists of replacing in the update rule \eqref{es:StochFB}, $\left(\bu_j\right)_{j \ge 1}$ by $\left(\overline{\bu}_j\right)_{j \ge 1}$ where, for every $j \ge 1$, $\overline{\bu}_j = \frac{1}{j} \sum_{i = 1}^j \bu_i$. The advantage is that it provides convergence guarantees for nondecaying stepsize sequences. Finally, the so-called \emph{composite mirror descent} methods, introduced in \cite{DuchiColt10}, can be viewed as extended versions of the SFB algorithm where the proximity operator is computed with respect to a non Euclidean distance (typically, a Bregman divergence).

In the last few years, a great deal of effort has been made to modify SFB when \black{the proximity operator of} \black $g \circ \bD$ does not have a simple expression, but when \black{$g$} \black can be split into several terms whose proximity operators are explicit. We can mention the stochastic proximal averaging strategy from \cite{Zhong2014}, the stochastic 
\emph{alternating direction method of mutipliers} (ADMM) from \cite{Ouyang13,Suzuki13,Zhong14b} and the alternating block strategy from \cite{Xu_Y_2014_block_sgicno} suited to the case when $g \circ \bD$ is a separable function. 

Another active research area addresses the search for strategies to improve the convergence rate of SFB. Two main approaches can be
distinguished in the literature. The first, adopted for example in \cite{Hu09,Lan2012,Lin_Q_2014_comp-optim-appl_Sparsity_spgmco,Atchade_Y_2014_Stochastic_pgga}, relies on subspace acceleration. In such methods, usually reminiscent of Nesterov's acceleration techniques in the deterministic case, the convergence rate is improved by using information from previous iterates for the construction of the new estimate. Another efficient way to accelerate the convergence of SFB is to incorporate in the update rule second-order information one may have on the cost functions. For instance, the method described in \cite{Duchi11adagrad} incorporates quasi-Newton metrics into the SFB and RDA algorithms, and the \emph{natural gradient} method from \cite{Amari98} can be viewed as a preconditioned SGD algorithm. The two strategies can be combined, as for example, in \cite{ChouzenouxS3MG}. 

\subsubsection{Adaptive filtering methods}
In adaptive filtering, stochastic gradient-like methods have been quite popular for a long time \cite{Widrow_B_1985_Adaptive_sp,Kushner_H_2003_stochastic_araa}. In this field, the functions $(\varphi_j)_{j \ge 1}$ often reduce to a least squares criterion
\begin{equation}
(\forall j \ge 1) \quad \varphi_j(\bh_j^T \bx,y_j) = (\bh_j^T \bx- y_j)^2
\end{equation}
where $\bx$ is the unknown impulse response.
However, a specific difficulty to be addressed is that the designed algorithms must be able to deal with dynamical problems the optimal solution of which
may be time-varying due to some changes in the statistics of the available data. 
In this context, it may be useful to adopt a multivariate formulation by
imposing, at each iteration $j \ge Q$
\begin{equation}
\by_j \simeq \bH_j \bx
\end{equation} 
where $\by_j = [y_j,\ldots,y_{j-Q+1}]^T$, $\bH_j = [\bh_j,\ldots,\bh_{j-Q+1}]^T$,
and $Q \ge 1$. 
\ModifJCP{This technique, reminiscent of mini-batch procedures in machine learning, constitutes the principle of \emph{affine projection algorithms}, the purpose of which is to accelerate the convergence speed}
\cite{Gay_S_1995_p-icassp_Fast_apa}.
Our focus now switches to recent work which aims to impose some sparse
structure on the desired solution. 

A simple method for imposing sparsity is to introduce a suitable adaptive preconditioning strategy in the stochastic gradient iteration,
leading to the so-called \emph{proportionate least mean square} method \cite{Khong_A_2006_p-asilomar_Efficient_usaf,Paleologu_C_2010_p-icassp_Improved_pNLMS},
which can be combined with affine projection techniques \cite{Hoshuyama_O_2004_p-icassp_Generalized_pvssa,Paleologu_C_2012_p-icassp_Regularization_ipapa}.
Similarly to the work already mentioned that has been developed in the machine learning community,
a second approach proceeds by minimizing penalized criteria such as \eqref{e:probminstochgen} where $g$ is a sparsity measure and $\bD = \boldsymbol{I}_N$.
In \cite{Chen_Y_2009_p-icassp_Sparse_lms,Chen_Y_2010_unpublished_Regularized_LMSa}, \emph{zero-attracting} algorithms are developed which are based on the stochastic 
subgradient method. These algorithms have been further extended to affine projection techniques in \cite{Meng_R_2011_procssdp_sparsity_aapaasi,Markus_L_2013_p-icassp_Affine_passi,Markus_L_2014_ieee-tsp_sparsity_adsaf}. Proximal methods have also been proposed in the context of adaptive filtering, grounded on the use of
 the forward-backward algorithm 
 \cite{Murakami2010}, 
 an accelerated version of it \cite{Yamagishi_M_2011_p-icassp_Acceleration_apfbs}, or primal-dual approaches \cite{Ono_S_2013_p-icassp_sparse_siu}. 
 It is interesting to note that \emph{proportionate affine projection algorithms} can be viewed as special cases of these methods \cite{Murakami2010}.
 Other types of algorithms have been proposed which provide extensions of the \black{\emph{recursive least squares}} \black method, which is known for its fast convergence 
 properties \cite{Angelosante2010,Babadi_B_2010-ieee-tsp_SPARSELS_srlsa,ChouzenouxS3MG}.
 Instead of minimizing a sparsity promoting criterion, it is also possible to formulate the problem as a feasibility problem where, at iteration $j\ge Q$, one searches for a vector
 $\bx$ satisfying both  $\sup_{j \le i \le j-Q+1} |y_i - \bh_i^T \bx| \le \eta$ and $\|\bx \|_1 \le \rho$, where $\| \cdot\|_1$ denotes the (possibly weighted) $\ell_1$ norm
 and $(\eta,\rho)\in ]0,+\infty[^2$.
Over-relaxed projection algorithms allow such kind of problems to be solved efficiently \cite{Kopsinis_Y_2011_ieee-tsp_online_ssisr,Slavakis_K_2013_ieee-tsp_Generalized_tosal}.

\subsection{Stochastic algorithms for solving deterministic optimization problems}\label{se:stochalgpbdet}
We now consider the deterministic optimization problem defined by \eqref{e:optMAp} and \eqref{e:seph}.
Of particular interest is the case when the dimensions $N$ and/or $M$ are very large \ModifJCP{(for instance, in \cite{Mairal14}, $M = 2500000$ and in \cite{Repetti_eusipco15}, $N = 100250$)}.

\subsubsection{Incremental gradient algorithms}

Let us start with incremental methods, which are dedicated to the solution of \eqref{e:optMAp} when $M$ is large, so that one prefers to exploit at each iteration a single term $\varphi_j$, usually through its gradient, rather than the global function $\varphi$.    
There are many variants of incremental algorithms, which differ in the assumptions made on the functions involved, on the stepsize sequence, and on the way of activating the functions
$(\varphi_i)_{1\le i\le  M}$. This order could follow either a deterministic \cite{Blatt2007} or a randomized rule. However, it should be noted that the use of randomization in the selection of the components presents some benefits in terms of convergence rates \cite{Bertsekas_D_P_2010_Incremental_gspmcos} which are of particular interest in the context of machine learning \cite{Schmidt2013,Defazio14}, where the user can only afford few full passes over the data. Among randomized incremental methods, the SAGA algorithm \cite{Defazio14b}, presented below, allows the problem defined in \eqref{e:optMAp} to be solved when the function $g$ is not necessarily smooth, by making use of the proximity operator introduced previously.
The $n$-th iteration of SAGA reads as
%
\begin{equation}
\begin{array}{l}
\bu_{n}   = \bh_{j_n}  \nabla \varphi_{j_n}(\bh_{j_n}^T \bx_n,y_{j_n}) - \bh_{j_n} \nabla \varphi_{j_n}(\bh_{j_n}^T \bz_{j_n,n},y_{j_n}) \\
\qquad \qquad + \frac{1}{M} \sum_{i=1}^M \bh_i \nabla \varphi_i(\bh_i^T \bz_{i,n},y_i)\\
\bz_{j_n,n+1} = \bx_n\\
\bz_{i,n+1} = \bz_{i,n} \quad \forall\, i \in \left\{1,\ldots,M\right\} \setminus \left\{j_n\right\} \\\
\bx_{n+1} = \prox_{\gamma g \circ \bD}(\bx_n - \gamma \bu_{n})
\end{array}
\label{eq:SAGA}
\end{equation}
where $\gamma \in ]0,+ \infty[$, for all $i \black{\in} \black \{1,\ldots,M\}$, $\bz_{i,1} = \bx_1\in \RR^N$, and $j_n$ is drawn from an i.i.d. uniform distribution on $\{1,\ldots,M\}$.
Note that, although the storage of the variables $\left(\bz_{i,n}\right)_{1 \leq i \leq M}$ can be avoided in this method, it is necessary to store the $M$ gradient vectors $\big(\bh_i \nabla \varphi_i(\bh_i^T \bz_{i,n},y_i)\big)_{1 \leq i \leq M}$. The convergence of Algorithm~\eqref{eq:SAGA} has been analyzed in \cite{Defazio14b}.
If the functions $(\varphi_i)_{1\le \i \le M}$ are $\beta^{-1}$-Lipschitz differentiable and $\mu$-strongly convex with $(\beta,\mu)\in ]0,+ \infty[^2$ and 
the stepsize $\gamma$ equals $\beta / (2(\mu\beta M + 1))$, then $(\Ex{ \{\| \bx_n - \overline{\bx}\|^2}\})_{n\in \mathbb{N}}$ goes to zero geometrically with rate $1 - \gamma$, where $\overline{\bx}$ is the solution to \black Problem~\eqref{e:optMAp}. When only convexity is assumed, a weaker convergence result is available.

The relationship between Algorithm~\eqref{eq:SAGA} and other stochastic incremental methods existing in the literature is worthy of comment. The main distinction arises in the way of building the gradient estimates $(\bu_n)_{n \geq 1}$. The standard incremental gradient algorithm, analyzed for instance in \cite{Bertsekas_D_P_2010_Incremental_gspmcos}, relies on simply defining, at iteration $n$, $\bu_n = \bh_{j_n}  \nabla \varphi_{j_n}(\bh_{j_n}^T \bx_n,y_{j_n})$. However, this approach, while leading to a smaller computational complexity per iteration and to a lower memory requirement, gives rise to suboptimal convergence rates \cite{Bertsekas_D_P_2010_Incremental_gspmcos,Nemirovski08}, mainly due to the fact that its convergence requires a stepsize sequence $(\gamma_n)_{n \geq 1}$ decaying to zero. Motivated by this observation, much recent work~\cite{Johnson2013,Schmidt2013,Konecny2014,Mairal14,Defazio14,Defazio14b} has been dedicated to the development of \emph{fast incremental gradient methods}, which would benefit from the same convergence rates as batch optimization methods, while using a randomized incremental approach. A first class of methods relies on a \emph{variance reduction approach} \cite{Johnson2013,Schmidt2013,Konecny2014,Defazio14b} which aims at diminishing the variance in successive estimates $(\bu_n)_{n \geq 1}$. All of the aforementioned algorithms are based on iterations which are similar to \eqref{eq:SAGA}. In the \emph{stochastic variance reduction gradient} method and the \emph{semi-stochastic gradient descent} method proposed in \cite{Johnson2013,Konecny2014}, a full gradient step is made at every $K$ iterations, $K \ge 1$, so that 
a single vector $\widetilde{\bz}_n$ is used instead of $\left(\bz_{i,n}\right)_{1 \leq i \leq M}$ in the update rule. This so-called mini-batch strategy leads to a reduced memory requirement at the expense of more gradient evaluations. As pointed out in \cite{Defazio14b}, the choice between one strategy or another may depend on the problem size and on the computer architecture. In the \emph{stochastic average gradient} algorithm (SAGA) from \cite{Schmidt2013}, a multiplicative factor $1/M$ is placed in front of the gradient differences, leading to a lower variance counterbalanced by a bias in the gradient estimates. It should be emphasized that the work in \cite{Johnson2013,Schmidt2013} is limited to the case when $g \equiv 0$. A second class of methods, closely related to SAGA, consists of applying  the proximal step to $\overline{\bz}_n - \gamma \bu_n$, where $\overline{\bz}_n$ is the average of the variables $\left(\bz_{i,n}\right)_{1 \leq i \leq M}$ (which thus need to be stored). This approach is retained for instance in the Finito algorithm \cite{Defazio14}
as well as in some instances of the \emph{minimization by incremental surrogate optimization} (MISO) algorithm, proposed in \cite{Mairal14}. These methods are of particular interest when the extra storage cost is negligible with respect to the high computational cost of the gradients. Note that the MISO algorithm relying on the majoration-minimization framework employs a more generic update rule than Finito 
and has proven convergence guarantees even when $g$ is nonzero. 

\subsubsection{Block coordinate approaches}
In the spirit of the Gauss-Seidel method, an efficient approach for dealing with Problem~\eqref{e:optMAp} when $N$ is large consists of resorting to block coordinate alternating strategies. Sometimes, such a block alternation can be performed in a deterministic manner \cite{TsengBCD2001,ChouzenouxBCVMFB2014}. However,
many optimization methods are based on fixed point algorithms, and it can be shown that with deterministic block coordinate strategies,
the contraction properties which are required to guarantee the convergence of such algorithms are generally no longer satisfied. In turn, by resorting to
stochastic techniques, these properties can be retrieved in some probabilistic sense \cite{Combettes_P_2014_Stochastic}. In addition, using stochastic rules
for activating the different blocks of variables often turns out to be more flexible.

To illustrate why there is interest in block coordinate approaches, let us split the target variable $\bx$ as
$[\bx_1^T,\ldots,\bx_K^T]^T$,
where, for every \black{$k \in \{1,\ldots,K\}$}, \black $\bx_k \in \RR^{N_k}$ is the $k$-th block of variables
with reduced dimension $N_k$ (with $N_1+\cdots+N_K = N$). Let us further assume that the regularization
function can be blockwise decomposed as
\begin{equation}
g(\bD \bx ) = \sum_{k=1}^K g_{1,k}(\bx_k)+g_{2,k}(\bD_k \bx_k)
\end{equation}
where, for every  \black{$k \in \{1,\ldots,K\}$}, \black $\bD_k$ is a matrix in $\RR^{P_k \times N_k}$, and
$g_{1,k}\colon\RR^{N_k}\to ]-\infty,+\infty]$ and $g_{2,k}\colon \RR^{P_k} \to ]-\infty,+\infty]$
are proper lower-semicontinuous  convex functions. Then, the stochastic
primal-dual proximal algorithm allowing us to solve Problem \eqref{e:optMAp} is given by
\begin{algorithm}[H]
\caption{Stochastic primal-dual proximal algorithm}
\begin{algorithmic}
\FOR {$n=1,2,\ldots$}
\FOR {$k=1$ to $K$}
\STATE \textbf{with probability} $\varepsilon_{k} \in (0,1]$ \textbf{do}
\STATE $\;\;\bv_{k,n+1} = (\operatorname{Id}-\prox_{\tau^{-1} g_{2,k}})(\bv_{k,n}+ \bD_k \bx_{k,n})$
\STATE $\;\;\bx_{k,n+1} = \prox_{\gamma g_{1,k}}\Big(\bx_{k,n}-\gamma \big(\tau\bD_k^T (2\bv_{k+1,n}-\bv_{k,n})$
\STATE $\qquad\qquad\qquad+\frac{1}{M} \sum_{i=1}^M \bh_{i,k} \nabla \varphi_i(\sum_{k'=1}^K \bh_{i,k'}^T \bx_{k',n},y_i)\big)\Big)$
\STATE \textbf{otherwise}
\STATE $\;\;\bv_{k,n+1} = \bv_{k,n},\;\bx_{k,n+1} = \bx_{k,n}.$
\ENDFOR
\ENDFOR
\end{algorithmic}
\label{algo:PDstoch}
\end{algorithm}
In the algorithm above, for every \black{$i \in \{1,\ldots,M\}$}, \black the scalar product $\bh_i^T \bx$ has
been rewritten in a blockwise manner as $\sum_{k'=1}^K \bh_{i,k'}^T \bx_{k'}$. 
Under some stability conditions on the choice of the positive step sizes $\tau$ and $\gamma$,
 $\bx_n = [\bx_{1,n}^T,\ldots,\bx_{K,n}^T]^T$ converges almost surely to a solution
 of the minimization problem, as $n\to + \infty$ (see \cite{Pesquet_J-C_jnca_Class_rpdado} for more technical details). 
It is important to note that the convergence result was established for arbitrary probabilities $\boldsymbol{\varepsilon} = [\varepsilon_{1},\ldots,\varepsilon_{K}]^T$, provided that the block activation probabilities $\varepsilon_{k}$ are positive and independent of $n$.
\black{Note that the} \black \ModifJCP{various blocks can also be activated in a dependent manner at a given iteration $n$.}
Like its deterministic counterparts (see \cite{Komodakis_N_2014_playing_dor} and the references therein), this algorithm enjoys the property of not requiring any matrix inversion, 
 which is of paramount importance when the matrices $(\bD_k)_{1\le k \le K}$ are of large size and do not have some simple forms.
 
 When $g_{2,k} \equiv 0$, the random block coordinate forward-backward algorithm is recovered as an instance of Algorithm~\ref{algo:PDstoch}
 since the dual variables $(\bv_{k,n})_{1\le k\le K,n\in \mathbb{N}}$ can be set to 0 and the constant $\tau$ becomes useless. 
 An extensive literature exists on the latter algorithm and its variants. In particular, its almost sure convergence was established 
 in \cite{Combettes_P_2014_Stochastic} under general conditions, whereas some worst case convergence rates
 were derived in \cite{Richtarik_P_2014_math-prog_Iteration_crbcdmmcf,Necoara_I_2014_com-opt-Random-cdaopcoflcc,Lu_Z_2015_On-carbcdm,Fercoq_O_2014_Accelerated_ppcd,Qu_Z_2014_Coordinate-dasIac}. In addition, if $g_{1,k} \equiv 0$, the \emph{random block coordinate descent} algorithm is obtained \cite{Nesterov_Yu_2012_siopt_Efficient_cdmhsop}.
 
 When the objective function minimized in Problem  \eqref{e:optMAp} is strongly convex, the random block coordinate forward-backward algorithm can be applied
 to the dual problem, in a similar fashion to the dual forward-backward method used in the deterministic case \cite{Combettes_P_2010_j-svva_dualization_srp}. 
 This leads to so-called dual ascent strategies which have become quite popular in machine learning \cite{SS2013,SS2014,Jaggi2014,Qu_Z_2014_Randomized_dcaas}.
 
 Random block coordinate versions of other proximal algorithms such as the Douglas-Rachford algorithm and ADMM have also been proposed \cite{Iutzeler_F_2013_p-dcc_Asynchronous_douradmm,Combettes_P_2014_Stochastic}.
 Finally, it is worth emphasizing that asynchronous distributed algorithms can be deduced from 
 various randomly activated block coordinate methods \cite{Pesquet_J-C_jnca_Class_rpdado,Bianchi_P_2014_stochastic_cdpda}. As well as dual ascent methods,
 the latter algorithms can also be viewed  as incremental methods.

\section{Areas of intersection: optimization-within-MCMC and MCMC-driven optimization}\label{Sec5}
There are many important examples of the synergy between stochastic simulation and optimization, including global optimization by simulated annealing, stochastic EM algorithms, and adaptive MCMC samplers \cite{casella:robert:2004}. In this section we highlight some of the interesting new connections between modern simulation and optimization that we believe are particularly relevant for the SP community, and that we hope  will stimulate further research in this community.

\subsection{Riemannian manifold MALA and HMC}
Riemannian manifold MALA and HMC exploit differential geometry for the problem of specifying an appropriate proposal covariance matrix $\boldsymbol{\Sigma}$ that takes into account the geometry of the target density $\pi$ \cite{girolami:2011}. These new methods stem from the observation that specifying $\boldsymbol{\Sigma}$ is equivalent to formulating the Langevin or Hamiltonian dynamics in an Euclidean parameter space with inner product $\langle\bw, \boldsymbol{\Sigma}^{-1} \bx\rangle$. Riemannian methods advance this observation by considering a smoothly-varying position dependent matrix $\boldsymbol{\Sigma}(\bx)$, which arises naturally by formulating the dynamics in a Riemannian manifold. The choice of $\boldsymbol{\Sigma}(\bx)$ then becomes the more familiar problem of specifying a metric or distance for the parameter space \cite{girolami:2011}. Notice that the Riemannian and the canonical Euclidean gradients are related by $\tilde{\nabla} g(\bx) = \boldsymbol{\Sigma}(\bx) \nabla g(\bx)$. Therefore this problem is also closely related to gradient preconditioning in gradient descent optimization discussed in Sec IV.B. Standard choices for $\boldsymbol{\Sigma}$ include for example the inverse Hessian matrix \cite{ zhang:2011, betancourt:2013}, which is closely related to Newton's optimization method, and the inverse Fisher information matrix \cite{girolami:2011}, which is the ``natural'' metric from an information geometry viewpoint and is also related to optimization by natural gradient descent \cite{Amari98}. These strategies have originated in the computational statistics community, and perform well in inference problems that are not too high-dimensional. Therefore, the challenge is to design new metrics that are appropriate for SP statistical models (see \cite{allassonniere:2012, marnissi:2014} for recent work in this direction).
\subsection{Proximal MCMC algorithms}
Most high-dimensional MCMC algorithms rely particularly strongly on differential analysis to navigate vast parameter spaces efficieoptimizationntly. Conversely, the potential of convex calculus for MCMC simulation remains largely unexplored. This is in sharp contrast with modern high-dimensional optimization described in Section \ref{Sec4}, where convex calculus in general, and proximity operators \cite{Mor62b,Combettes_2010} in particular, are used extensively. This raises the question as to whether convex calculus and proximity operators can also be useful for stochastic simulation, especially for high-dimensional target densities that are log-concave, and possibly not continuously differentiable.

This question was studied recently in \cite{pereyra:2015} in the context of Langevin algorithms. As explained in Section II.B, Langevin MCMC algorithms are derived from discrete-time approximations of the time-continuous Langevin diffusion process \eqref{diffusion}. Of course, the stability and accuracy of the discrete approximations determine the theoretical and practical convergence properties of the MCMC algorithms they underpin. The approximations commonly used in the literature are generally well-behaved and lead to powerful MCMC methods.  \black{However, they can perform poorly if $\pi$ is not sufficiently regular, for example if $\pi$ is not continuously differentiable, if it is heavy-tailed, or if it has lighter tails than a Gaussian distribution. This drawback limits the application of MCMC approaches to many SP problems, which rely increasingly on models that are not continuously differentiable or that involve constraints.}\black

Using proximity operators, the following proximal approximation for the Langevin diffusion process \eqref{diffusion} was recently proposed in \cite{pereyra:2015}
\begin{equation}\label{PULA}
X^{(t+1)} \sim \mathcal{N}\left(\prox_{\frac{-\delta}{2} \log\pi} \left(X^{(t)}\right),\delta\mathbb{I}_N\right)
\end{equation}
as an alternative to the standard forward Euler approximation $X^{(t+1)} \sim \mathcal{N}\left(X^{(t)} + \frac{\delta}{2}\nabla\log\pi\left(X^{(t)}\right),\delta\mathbb{I}_n\right)$ used in MALA\footnote{Recall that $\prox_{\varphi}(\bv)$ denotes the proximity operator of $\varphi$ evaluated at $\bv\in \RR^N$ \cite{Mor62b,Combettes_2010}.}. Similarly to MALA, the time step $\delta$ can be adjusted online to achieve an acceptance probability of approximately $50\%$. It was established in \cite{pereyra:2015} that when $\pi$ is log-concave, \eqref{PULA} defines a remarkably stable discretization of \eqref{diffusion} with optimal theoretical convergence properties. Moreover, the ``proximal'' MALA resulting from combining \eqref{PULA} with an MH step has very good geometric ergodicity properties. In \cite{pereyra:2015}, the algorithm efficiency was demonstrated empirically on challenging models that are not well addressed by other MALA or HMC methodologies, including an image resolution enhancement model with a total-variation prior. Further practical assessments of proximal MALA algorithms would therefore be a welcome area of research.

Proximity operators have also been used recently in \cite{chaari:2014} for HMC sampling from log-concave densities that are not continuously differentiable. The experiments reported in \cite{chaari:2014} show that this approach can be very efficient, in particular for  SP models involving high-dimensionality and non-smooth priors. Unfortunately, theoretically analyzing HMC methods is difficult, and the precise theoretical convergence properties of this algorithm are not yet fully understood. We hope future work will focus on this topic.

\subsection{optimization-driven Gaussian simulation}
The standard approach for simulating from a multivariate Gaussian distribution with precision matrix $\boldsymbol{Q} \in \mathbb{R}^{n\times n}$ is to perform a Cholesky factorization $\boldsymbol{Q} = \boldsymbol{L}^T \boldsymbol{L}$, generate an auxiliary Gaussian vector $\bw \sim \mathcal{N}(0,\mathbb{I}_N)$, and then obtain the desired sample $\bx$ by solving the linear system $\boldsymbol{L}\bx = \bw$ \cite{rue:2001}. The computational complexity of this approach generally scales at a prohibitive rate $\mathcal{O}(N^3)$ with the model dimension $N$, making it impractical for large problems, (note however that there are specific cases with lower complexity, for instance when $\boldsymbol{Q}$ is Toeplitz \cite{trench:1964}, circulant \cite{geman:1995} or sparse \cite{rue:2001}). 

optimization-driven Gaussian simulators arise from the observation that the samples can also be obtained by minimizing a carefully designed stochastic cost function \cite{lafferty:2010,orieux:2012}. For illustration, consider a Bayesian model with Gaussian likelihood $\by|\bx \sim \mathcal{N}(\boldsymbol{H}\bx,\boldsymbol{\Sigma}_{\by})$ and Gaussian prior $\bx \sim \mathcal{N}(\bx_0,\boldsymbol{\Sigma}_{\bx})$, for some linear observation operator $\boldsymbol{H} \in \mathbb{R}^{N \times M}$, prior mean $\bx_0 \in \mathbb{R}^N$, and positive definite covariance matrices $\boldsymbol{\Sigma}_{\bx} \in \mathbb{R}^{N \times N}$ and $\boldsymbol{\Sigma}_{\by} \in \mathbb{R}^{M \times M}$. The posterior distribution $p(\bx|\by)$ is Gaussian with mean $\boldsymbol{\mu} \in \mathbb{R}^{N}$ and precision matrix $\boldsymbol{Q} \in \mathbb{R}^{N \times N}$ given by
$$
\boldsymbol{Q} = \boldsymbol{H}^T \boldsymbol{\Sigma}_{\by}^{-1} \boldsymbol{H} + \boldsymbol{\Sigma}_{\bx}^{-1}
$$
$$
\boldsymbol{\mu} = \boldsymbol{Q}^{-1}\left(\boldsymbol{H}^T \boldsymbol{\Sigma}_{\by}^{-1}\by + \boldsymbol{\Sigma}_{\bx}^{-1}\bx_0\right).
$$
Simulating samples $\bx | \by \sim \mathcal{N}(\boldsymbol{\mu},\boldsymbol{Q}^{-1})$ by Cholesky factorization of $\boldsymbol{Q}$ can be computationally expensive when $N$ is large. Instead, optimization-driven simulators generate samples by solving the following ``random'' minimization problem
\begin{equation}\label{PO}
\begin{split}
\bx = \argmin_{\bu \in \mathbb{R}^N} \,\,&\left(\bw_1 - \boldsymbol{H}\bu\right)^T \boldsymbol{\Sigma}_{\by}^{-1} \left(\bw_1 - \boldsymbol{H}\bu\right)\\ &\,+ \left(\bw_2 - \bu\right)^T \boldsymbol{\Sigma}_{\bx}^{-1} \left(\bw_2 - \bu\right)
\end{split}
\end{equation}
with random vectors $\bw_1 \sim \mathcal{N}(\by, \boldsymbol{\Sigma}_y)$ and $\bw_2 \sim \mathcal{N}(\bx_0, \boldsymbol{\Sigma}_x)$. It is easy to check that if \eqref{PO} is solved exactly, then $\bx$ is a sample from the desired posterior distribution $p(\bx|\by)$. From a computational viewpoint, however, it is significantly more efficient to solve \eqref{PO} approximately, for example by using a few linear conjugate gradient iterations \cite{orieux:2012}. The approximation error can then be corrected by using an MH step \cite{gilavert:2015},
at the expense of introducing some correlation between the samples and therefore reducing the total effective sample size. Fortunately, there is an elegant strategy to determine automatically the optimal number of conjugate gradient iterations that maximizes the overall efficiency of the algorithm \cite{gilavert:2015}. 


%

\section{Conclusions and Observations}\label{Sec6}

In writing this paper we have sought to provide an introduction to stochastic simulation and optimization methods in a tutorial format, but which also raised some interesting topics for future research. We have addressed a variety of MCMC methods and discussed surrogate methods, such as variational Bayes, the Bethe approach, belief and expectation propagation, and approximate message passing. We also discussed a range of recent advances in optimization methods that have been proposed to solve stochastic problems, as well as stochastic methods for deterministic optimization. Subsequently, we highlighted new methods that combine simulation and optimization, such as proximal MCMC algorithms and optimization-driven Gaussian simulators. Our expectation is that future
methodologies will become more flexible. Our community has successfully applied computational inference methods, as we have described, to a plethora of challenges across an enormous range of application domains. Each problem offers different challenges, ranging from model dimensionality and complexity, data (too much or too little), inferences, accuracy and computation times. Consequently, it seems not unreasonable to speculate that the different
computational methodologies discussed in this paper will evolve to become more adaptable, with their boundaries becoming less well defined, and with the development of algorithms that make use of simulation, variational approximations
and optimization simultaneously. Such an approach is more likely to be able to handle an even wider range of models, datasets, inferences, accuracies and computing times in a computationally efficient way.

\footnotesize
\bibliographystyle{IEEEtran}
\bibliography{phil,bibtutorial,bibliographySec2}
\end{document}